\documentclass[fleqn,usenatbib]{mnras}

\usepackage{newtxtext,newtxmath}

\usepackage[T1]{fontenc}

\DeclareRobustCommand{\VAN}[3]{#2}
\let\VANthebibliography\thebibliography
\def\thebibliography{\DeclareRobustCommand{\VAN}[3]{##3}\VANthebibliography}

\usepackage{graphicx}	
\usepackage{amsmath}	
\usepackage{CJKutf8}

\usepackage{soul}
\defcitealias{rulePlungingRegionThin2025a}{R25}
\defcitealias{mummeryAccretionInnermostStable2023}{MB23}

\title[Plunging region across the spin range]{The plunging region of thin accretion discs across the black hole spin range}

\author [Jake Rule et al.]{Jake Rule$^{1}$\thanks{E-mail:
jake.rule@physics.ox.ac.uk}, Andrew Mummery$^{2,3}$, Steven Balbus$^{1,3}$, James M. Stone$^2$, Lizhong Zhang (张力中)$^{2,4}$\\
$^1$Oxford Astrophysics, Denys Wilkinson Building,  Keble Road, Oxford, OX1 3RH, United Kingdom \\
$^2$School of Natural Sciences, Institute for Advanced Study, 1 Einstein Drive, Princeton, NJ 08540, USA \\
$^3$Oxford Theoretical Physics, Beecroft Building,  Clarendon Laboratory, Parks Road, Oxford, OX1 3PU, United Kingdom \\
$^4$Center for Computational Astrophysics, Flatiron Institute, New York, NY 10010, USA
}

\date{Accepted XXX. Received YYY; in original form ZZZ}

\pubyear{\the\year{}}

\begin{document}
\begin{CJK}{UTF8}{gbsn}
\label{firstpage}
\pagerange{\pageref{firstpage}--\pageref{lastpage}}
\maketitle
\end{CJK}
\begin{abstract}
We compute and test analytic models for the plunging region dynamics, thermodynamics, and magnetic fields against dedicated 3D global general relativistic magnetohydrodynamics (MHD) simulations of thin accretion discs around black holes across the spin range, using the code {\tt ATHENAK}. We find that the dynamics of the plunging fluid closely resembles that of a gravity-dominated geodesic plunge, with the best agreement at low spins. Additionally, we find good agreement between the thermodynamic framework and the simulated quantities across the spin range. Finally, we develop a new model for the magnetic fields in the plunging region that assumes a fixed geodesic inflow, into which the magnetic fields are frozen. Overall, our simulations are in good concordance with this model, albeit with some discrepancies that suggest a degree of non-ideal MHD dissipation. In addition, we investigate how the MHD stresses in the plunging region depend on the black hole spin, interpreting our results through the lens of our flux-freezing model. We find that the magnitude of the stress increases as the black hole spin is increased in the prograde direction. This question is of particular importance for observers who wish to determine the black hole spin from X-ray measurements of the inner accretion disc, since a low-stress, high-spin solution is degenerate with a high-stress, low-spin solution. The spin-stress relationship that we report is approximately orthogonal to the contour of degenerate spin-stress pairings, indicating that the degeneracy is not fundamental. We show this explicitly for the case of M33 X-7.
\end{abstract}




\section{Introduction}
Accreting black holes emit copious amounts of radiation, especially from their innermost regions. It is only by probing the regions close to a black hole that we are able to constrain its angular momentum, $J_\mathrm{bh}$. The angular momentum is generally normalised to a length via $a=J_\mathrm{bh}/Mc$, where $M$ is the black hole mass and $c$ the speed of light.  If $a$ is further reduced by expressing it in units of the gravitational radius $r_g=GM/c^2$ ($G$ is the gravitational constant), then general relativity restricts its dimensionless value to $-1\leq a\leq 1$.
\par
The $a$ parameter is one of two parameters that fully describe an astrophysical black hole: the other is the mass $M$. This is the classical `No-hair theorem'. Accurate measurements of the spin of black holes are crucial to a good understanding of the formation and growth history of black holes across the mass range from the stellar mass black holes found in X-ray Binary Systems (XRBs) to the supermassive black holes at the centre of Tidal Disruption Events (TDEs) and Active Galactic Nuclei (AGNs).
\par
An important feature of the spacetime around a black hole is the existence of an Innermost Stable Circular Orbit (ISCO). Interior to the radial location of this orbit, accreting matter follows a plunging trajectory directly into the horizon of the black hole, unless there is intervening thermal or magnetic support. The region between the ISCO and the event horizon is therefore known as the plunging region. The location of the ISCO is strongly spin-dependent \citep[e.g.][]{bardeenRotatingBlackHoles1972}. Since the ISCO fundamentally changes the nature of the flow of the accreting material, any change in its position leaves a spin-dependent imprint on the radiation emitted by the accreting material. Measuring this imprint via the emitted radiation from the flow allows one, in principle, to determine the spin of the black hole.
\par
In practice this is difficult to do, as it requires detailed modelling of both the hot accreting material and the emission (and propagation) of the radiation from disc to observer. Continuum fitting methods model the thermal emission from the main body of the accretion disc as a series of annuli locally radiating as a colour-corrected black body \citep[][]{zhangBlackHoleSpin1997, mcclintockBlackHoleSpin2014, reynoldsObservationalConstraintsBlack2021}. This approach requires a determination of the effective temperature of the disc as a function of radius. Typically, the \cite{novikovAstrophysicsBlackHoles1973} or \cite{pageDiskAccretionBlackHole1974} models are used for this, which are relativistic extensions of the \cite{shakuraBlackHolesBinary1973} physically thin disc model. These models all share a common idealisation: they assume a vanishing stress boundary condition at the ISCO, in effect truncating the disc by hand. There is now a consensus that analytical evidence \citep{gammieEfficiencyMagnetizedThin1999, krolikMagnetizedAccretionMarginally1999, agolMagneticStressMarginally2000}, numerical evidence \citep{shafeeThreeDimensionalSimulationsMagnetized2008,nobleDirectCalculationRadiative2009, nobleDependenceInnerAccretion2010, pennaSimulationsMagnetizedDiscs2010, schnittmanDiskEmissionMagnetohydrodynamic2016, dhangEnergyExtractionBlack2025, rulePlungingRegionThin2025a,lancovaRadiativeGRMHDSimulations2026}, and finally observational evidence \citep{mummeryContinuumEmissionPlunging2024,mummeryPlungingRegionEmission2024} all point to a finite stress at the ISCO, driven by MHD turbulence. This stress averts the sharp truncation of the disc at the ISCO, allowing material to flow through this region with a smoothly varying (and non-zero!) temperature. The passing material clearly must radiate, contributing significantly to the emission observed - especially at high energies.  Clearly, modelling the emission from material interior to the ISCO is essential to determine accurately the spin of the central black hole.
\par
Both the magnitude of the MHD stresses at the ISCO and the central black hole spin are free parameters to be extracted from observations, but there is some degeneracy in their determination  \citep{gammieEfficiencyMagnetizedThin1999, mummeryRapidBlackHole2025a}. Any dissipation caused by a large ISCO stress will heat the gas at the start of its plunge, which would reveal itself in the form of an excess of high-energy emission. This excess may be conflated with emission coming from comparable hot gas \emph{exterior} to the ISCO of a more rapidly spinning black hole with a lower ISCO stress, one in which the ISCO is closer to the event horizon and has fallen deeper into the potential well of the black hole.  To what extent these parameters are truly free (and hence both the existence and extent of the degeneracy) is, unfortunately, a complex question.  It requires full three-dimensional GRMHD simulations of both the plunging region and the main disc. Relatively little work has been done so far to study explicitly the relationship between the ISCO stress and the black hole spin. \cite{pennaSimulationsMagnetizedDiscs2010} presented a survey of thin-disc GRMHD simulations, some with varying black hole spin. However, they did not discuss any trend in the ISCO stress as a function of the black hole spin; rather, in their conclusions, they focused on the effects of the scale height. \cite{schnittmanDiskEmissionMagnetohydrodynamic2016} also surveyed a variety of black hole spins, but discussed primarily the simulated spectra that were produced in their GRMHD simulations.
\par
\cite{mummeryAccretionInnermostStable2023} (hereafter \citetalias{mummeryAccretionInnermostStable2023}) developed an analytic accretion model for the plunging fluid that may be used to model the thermodynamics that must underpin an emission model for the plunging material. The model has one parameter which may be related to the ISCO stress, and four key assumptions:
\begin{enumerate}
    \item Gravity is the dominant forcing term in the relativistic flow equations.
    \item The mass accretion rate, $\dot{M}$, is constant.
    \item The plunging flow is adiabatic.
    \item Vertical hydrostatic equilibrium is maintained.
\end{enumerate}
In \cite{rulePlungingRegionThin2025a} (hereafter \citetalias{rulePlungingRegionThin2025a}), we tested this model against a single, high-resolution, GRMHD simulation of a thin disc around a Schwarzschild black hole. The fluid followed geodesic plunging trajectories very closely, strongly suggesting that item (i) of the above list was an excellent approximation. The principal deviation from the original \citetalias{mummeryAccretionInnermostStable2023} model was item (iii): the calculated flow was not adiabatic, but actively heated. This may be caused by the release of magnetic energy from numerical dissipation of current sheets of oppositely directed field lines in the mid-plane of the plunging flow (the released energy is captured as heat in the code). Including an additional ad-hoc model in the \citetalias{mummeryAccretionInnermostStable2023} framework to account for this additional heating, good fits were found for the thermodynamic variables (density, pressure and temperature) as a function of the model parameter, suggesting that (ii) and (iv) were also valid. In \citetalias{rulePlungingRegionThin2025a}, we reported a $\delta_\mathcal{J} \approx 5.3 \%$ drop in the specific angular momentum ($U_\phi$) from the ISCO to the event horizon. Subsequent analysis has revealed that this was, in fact, an overestimate \footnote{See Section \ref{sec:GRMHDSims}.}. This error does not qualitatively affect the results of the thermodynamic fitting. The corrected value is $\delta_\mathcal{J} \approx 1.2 \%$. Since the order of magnitude is unchanged, the interpretation remains unchanged. The value is indicative of an ISCO stress large enough to produce measurable dissipation, averting any abrupt changes in the thermodynamic quantities at the ISCO, yet small enough that geodesic infall still represents an excellent approximation to the flow dynamics.
\par
To test the validity of the \citetalias{mummeryAccretionInnermostStable2023} framework as a robust method for spin measurement, it is important to investigate whether its predictions also apply to Kerr black holes more generally. It is also of considerable interest to illuminate any dependence that the ISCO stress {\em itself} may have on the black hole's spin parameter. These are the subjects addressed in this paper. We have produced a set of global three-dimensional GRMHD simulations that survey a range of central black hole spin parameters and are in all other respects as similar as possible to the zero-spin simulation carried out in \citetalias{rulePlungingRegionThin2025a}. 
\par
The structure of the paper is as follows. In \S  \ref{sec:GRMHDSims} we will set out the details of our simulation setup. In \S \ref{sec:Results} we will present the results of these simulations, exploring three key questions: How do the dynamics of the flow compare to geodesic inflow? How do the thermodynamics compare to the \citetalias{mummeryAccretionInnermostStable2023} (+ \citetalias{rulePlungingRegionThin2025a} non-adiabatic heating) solutions? How does the ISCO stress depend on the black hole spin? In \S \ref{sec:Flux-Freezing}, we will develop an analytic framework to add `frozen-in' magnetic fields to the existing \citetalias{mummeryAccretionInnermostStable2023} framework, enabling us to draw precise comparisons between the GRMHD simulations and analytic theory. In \S  \ref{sec:spinstressdegen}, we will discuss whether our results can help to break the observational spin-stress degeneracy. Finally, we will draw conclusions from our work in \S \ref{sec:conclusions}.
\section{GRMHD Simulations}
\label{sec:GRMHDSims}
In this paper, we simulate thin accretion discs across the black hole spin range within the ideal GRMHD framework, using the {\tt ATHENAK} code \citep[][]{stoneAthenaKPerformanceportableVersion2026}. We analyse our data in Spherical Kerr-Schild (SKS) coordinates ($t, r, \theta, \phi$), which are smoothly horizon-penetrating (unlike the well-known Boyer-Lindquist coordinates \cite{boyerMaximalAnalyticExtension1967}). Throughout, unless specified, we will adopt code units, in which we set $GM=c=1$, quote distances in units of $r_g=GM/c^2$ (which is unit radius), and times in units of $t_g=GM/c^3$.
\par
The purpose of this work is to survey a variety of black hole spin parameters.  These are $a/M \equiv cJ_\mathrm{bh}/GM^2 = \{-0.9,-0.5, 0.0,0.3,0.5,0.6,0.7,0.9\}$. The Schwarzschild simulation ($a/M=0.0$) is the same high-resolution simulation presented in \citetalias{rulePlungingRegionThin2025a}. For the purposes of computational feasibility, it was necessary to reduce the global resolution of the Kerr black hole simulations by a factor of two. This corresponds to a minimum cell spacing of $0.05 \, r_g \, \text{cell}^{-1}$. The mesh-refinement pattern and simulation domains are otherwise identical. We adopt a \cite{fishboneRelativisticFluidDisks1976} initial torus with the same inner radius of $r_\text{in}=10r_g$ and pressure maximum of $r_\text{max}=16 r_g$ for all simulations. It should be noted that due to the spin dependence of the equilibrium torus solution, the initial-time tori are not identical across simulations. In all cases, we normalise the density in code units so that the maximum density in the initial torus is $\rho_\text{max}=1$. However, where we compare between different simulations, we rescale all quantities so that the conserved mass in the initial torus matches that of the Schwarzschild simulation ($a/M=0.0$). We use the same 4-loop magnetic field configuration as \citetalias{rulePlungingRegionThin2025a}, with a minimum plasma beta parameter $\beta_\text{min}= \left(\max(u_g)/\max(u_\text{mag})\right)_\text{torus}=100$ for all simulations. This choice of magnetic initial condition places our simulations in the weakly magnetised (SANE) regime. The dimensionless magnetic flux parameter evaluated at the event horizon $\varphi_{\rm hor} \equiv \Phi_{\rm hor} (\dot{M}_{\rm hor} r_g^2 c)^{-1/2}$ is bounded by $0.1 \leq \varphi_{\rm hor}\leq 1$ for all of the simulations throughout their entire duration. For context, magnetically arrested discs (MADs) typically have a normalised flux of $\varphi_{\rm hor} \sim 50$ \citep[e.g.][]{narayanJetsMagneticallyArrested2022,scepiMagneticSupportWinddriven2024}. To define $\varphi_{\rm hor}$, we have invoked the hemispherical magnetic horizon flux $\Phi_{\rm hor}$ and the horizon mass accretion rate $\dot{M}_{\rm hor}$, which we define in the following way (see \citetalias{rulePlungingRegionThin2025a} or \cite{tchekhovskoyEfficientGenerationJets2011}):
\begin{align}
    \Phi_{\rm hor} = \left. \frac{1}{2} \int_0^\pi \int_0 ^{2 \pi} \vert B^r \vert \sqrt{\vert g \vert} {\rm d} \phi {\rm d} \theta \right \vert_{r=r_{\rm hor}} \, , \\
    \dot{M}_{\rm hor} = -\left. \int_0^\pi \int_0 ^{2 \pi} \rho U^r \sqrt{\vert g \vert} {\rm d} \phi {\rm d} \theta \right \vert_{r=r_{\rm hor}} \, .
\end{align}
\par
Finally, to keep the discs thin, we adopt exactly the same cooling function prescription as \citetalias{rulePlungingRegionThin2025a} \citep[first proposed by][]{pennaSimulationsMagnetizedDiscs2010}, with a target entropy constant of $K_{\rm t} = 0.001$. The cooling function is given by:
\begin{equation}
    \mathcal{L} = -u_g \frac{\ln(K/K_{\rm t})}{\tau_{\rm cool}},
\end{equation}

where $u_g$ is the internal energy of the gas and $K = P \rho^{-\gamma_{\rm ad}}$ is related to the specific entropy via $s = k_{\rm B} \ln(K)/(\gamma_{\rm ad} -1)$. Where $K<K_{\rm t}$, we set $\mathcal{L}=0$. Our cooling time-scale is $\tau_{\rm cool} = 2 \pi (a + R^{3/2} r_g^{-1/2})$, where $R = r \sin(\theta)$ is the cylindrical radius. See \citetalias{rulePlungingRegionThin2025a} for a more detailed discussion of this particular choice of cooling prescription and its specific implementation in {\tt ATHENAK}. Cooling ensures that all of the simulations settle to a thin scale-height aspect ratio of $ 0.04 \leq h/r \leq 0.07$ at $1.5\, r_I$.
\par
In \citetalias{rulePlungingRegionThin2025a}, we implemented a primitive state smoothing operator for `bad cells' \footnote{Those for which either C2P has failed or returned a thermal energy that is below the floor value.} to alleviate the unavoidable error in the conserved to primitive variable inversion process (C2P) for cells with a very large total energy and a much smaller thermal energy (either due to a large magnetic or inertial component). We have subsequently discovered that this process `smooths out' high mass flux cells (with a large inertial energy) that can carry significant mass and angular momentum, leading to unphysical non-conservation of these quantities. To compensate for this, we have now restricted the smoothing to just the thermal energy component of the primitive state for the vast majority of the affected cells. To implement this fix, we restarted all of the affected simulations and ran them for a suitable averaging window ($\Delta t = 5,000 \, t_g$).
\par
We will present averaged quantities calculated in exactly the same way as in \citetalias{rulePlungingRegionThin2025a}. Unless otherwise specified, we will average quantities over the full azimuthal domain, and from $\theta_u=76^\circ$ to $\theta_l=104^\circ$ in the vertical direction. This avoids including unbound, non-disc material in the polar regions. Our specific polar bounds are a practical choice: when converted from Cartesian to spherical coordinates, this wedge captures the majority of the most refined regions (see Table 1 of \citetalias{rulePlungingRegionThin2025a}), whilst excluding the unrefined regions above and below the equatorial plane. Finally, we integrate from $t_i = 20\,000 \, t_g$ to $t_f = 25\,000 \, t_g$ in the time domain, which is after the aforementioned restart.
\section{Results}
\label{sec:Results}
\subsection{Geodesic inflow}
\label{sec:geoinflow}
In our previous analysis of a disc around a Schwarzschild black hole, we found excellent agreement between the density-weighted average radial component of the 4-velocity and that of simple geodesic inflow solutions with the same energy and angular momentum as a particle on a circular orbit at the ISCO (see \citetalias{rulePlungingRegionThin2025a}). Despite the existence of a {\em thermally} important magnetic stress, we determined that it was not large enough to substantively alter the \emph{dynamics} of the fluid from that of geodesic infall, especially when close to the horizon. A testable prediction of the \cite{cunninghamEffectsRedshiftsFocusing1975,mummeryInspiralsInnermostStable2022} geodesic solution is that the quantity $\sqrt{r_I}\,U^r$ should have a universal, self-similar radial profile that is independent of the black hole spin,
\begin{equation}
    \sqrt{r_I}\,U^r = -c \sqrt{\frac{2r_g}{3}}\left(\frac{r_I}{r}-1\right)^{\frac{3}{2}}.
    \label{eq:urgeo}
\end{equation}
It is important to note that this geodesic solution cannot correspond to an exact accretion scenario on its own, as there is no inflow at the ISCO itself: it must reduce to a purely azimuthal circular orbit there by construction! In practice, to construct thermodynamic profiles, \citetalias{mummeryAccretionInnermostStable2023} insert a parametrised offset velocity ($U_I$) in addition to the pure geodesic solution. This is a convenient approximation in the trans-ISCO region that we expect will quickly relax to the pure geodesic solution as the flow approaches the horizon.
\begin{figure*}
    \centering
    \includegraphics[width=0.49\linewidth]{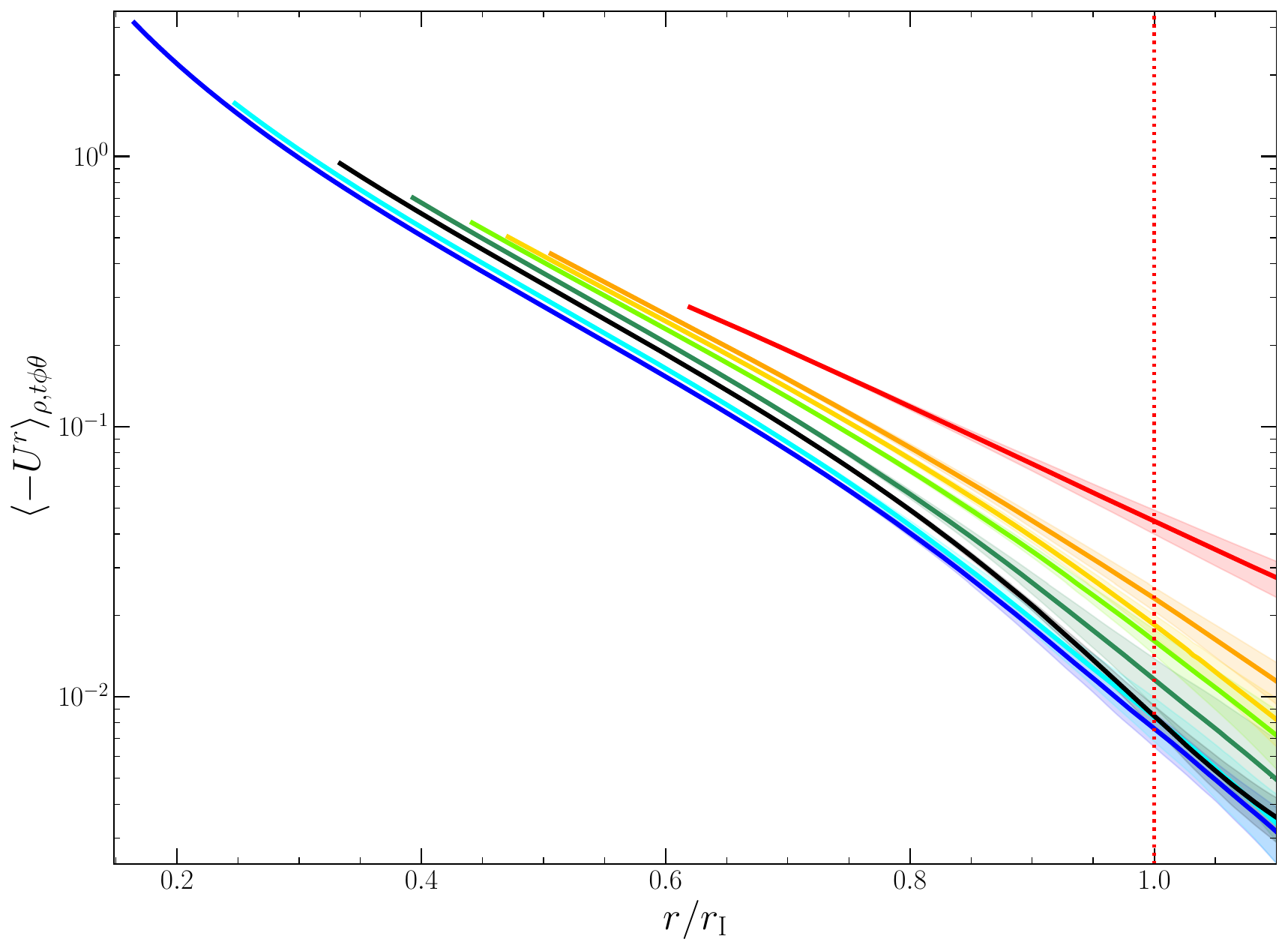}
    \includegraphics[width=0.49\linewidth]{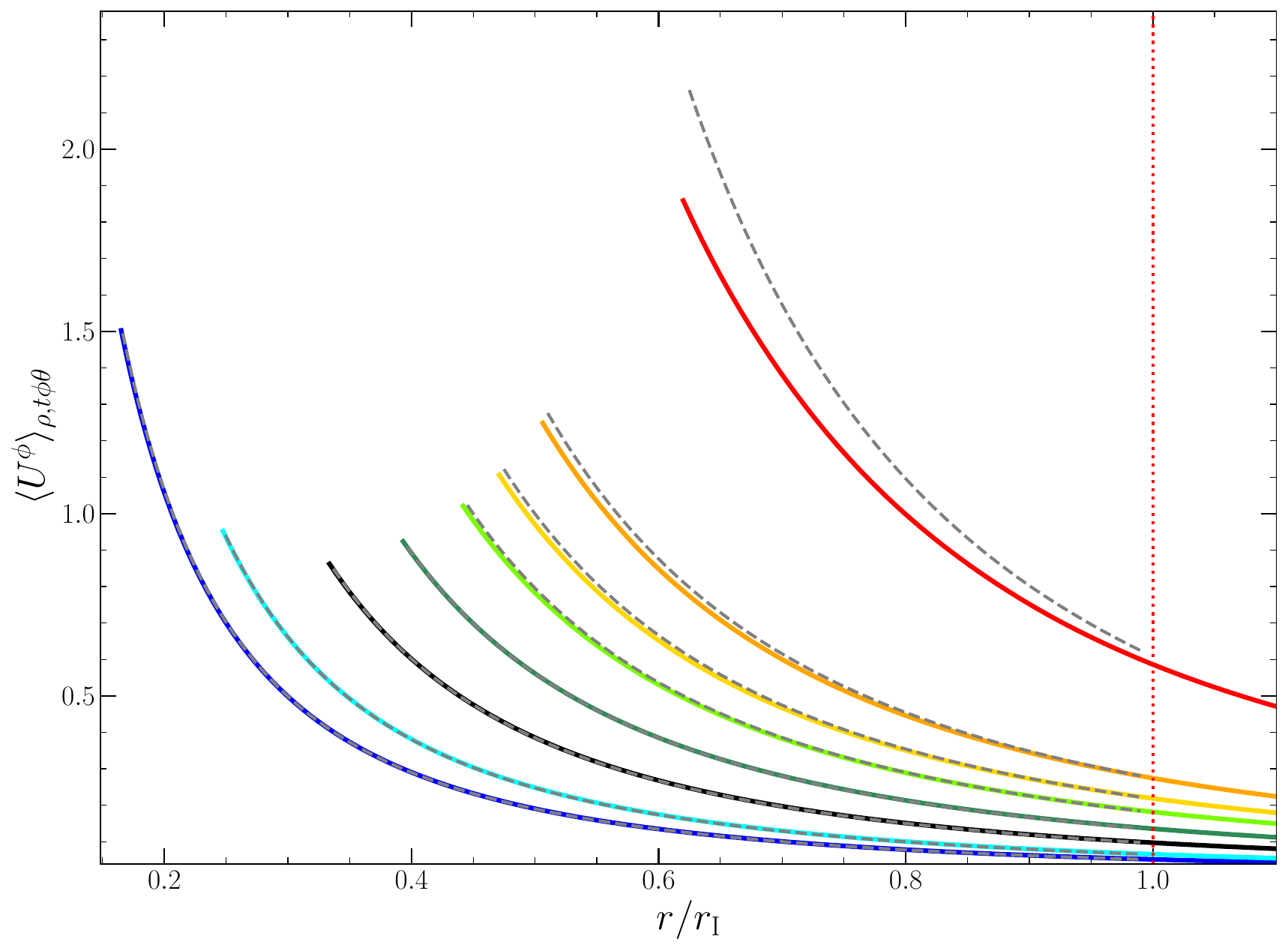}
    \includegraphics[width=0.6\linewidth]{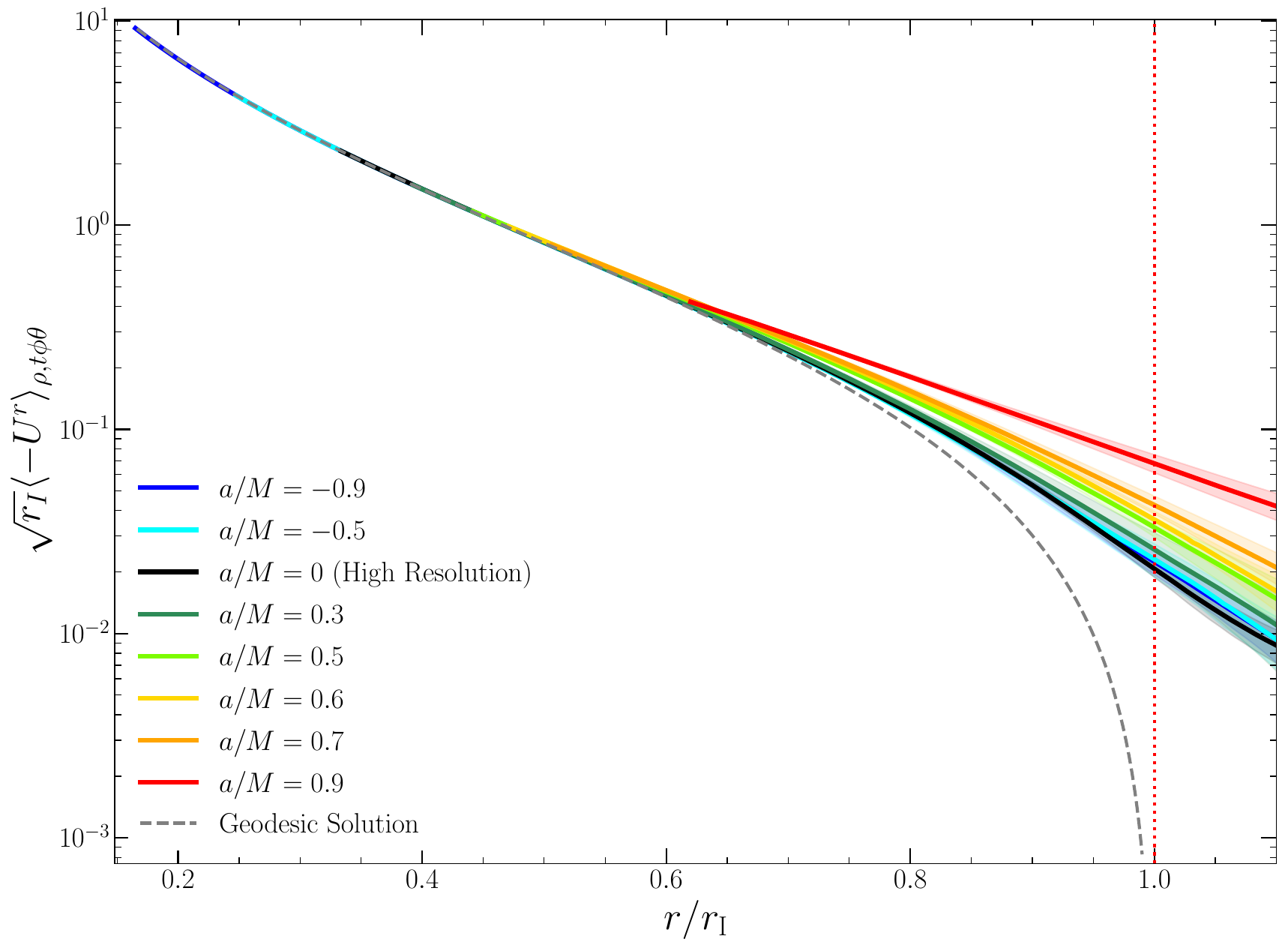}
    \caption{The radial 4-velocity ($U^r$), both unscaled (top-left) and scaled by $\sqrt{r_I}$ (bottom),  and the azimuthal 4-velocity ($U^\phi$) (top-right). The solid lines are the density-weighted quantities averaged over time, azimuth and the vertical direction, for each simulation. The dashed line in the bottom panel is the (spin independent) radial geodesic solution \citep[see][]{cunninghamEffectsRedshiftsFocusing1975,mummeryInspiralsInnermostStable2022}, whilst the dashed lines in the top-right panel are the (spin dependent) azimuthal geodesic solutions corresponding to the black hole spin of each simulation.}
    \label{fig:uprofiles}
\end{figure*}

\par
In the bottom panel of Fig.\,\ref{fig:uprofiles}, we compare the right-hand side of Eq.\,\ref{eq:urgeo} (the dashed line) to the density-weighted average profiles of $\sqrt{r_I}\,U^r$ for each simulation, which vary principally by their black hole spin. It is clear that for each simulation, the profile trends toward the same geodesic curve, irrespective of spin. For retrograde to moderate prograde spins, there is a clear overlap between the geodesic model and the simulation before the horizon is reached. For more rapid prograde spins, the plunging region is not physically large \footnote{Note that throughout this paper when discussing the `physical size' (or similar) of the plunging region, we refer to the proper time taken for a plunging fluid element on a (time-like) geodesic trajectory to fall from the ISCO to the horizon, as this is the relevant measure for our plunging inflow problem. One may verify that this measure is largest for $a/M=-1$ and smallest for $a/M = 1$.} enough for the profile to blend completely with the geodesic curve. However, it is noteworthy that even these profiles still trend \emph{towards} the geodesic profile near the horizon.
\par
In the top-left of Fig.\,\ref{fig:uprofiles}, we plot the unscaled $U^r$ profiles. There is a clear systematic increase in the radial crossing velocity at the ISCO ($U_I$) as a function of increasing prograde spin. This is to be expected. Since turbulent eddies that jump over the ISCO preferentially plunge inwards, we expect the scale of this crossing velocity to be set by one-sided turbulent fluctuations \citep{mummeryDynamicsAccretionFlows2024}. The scale of these turbulent fluctuations will be limited by the sound speed at the ISCO. As the prograde spin is increased, the ISCO moves inwards and deeper into the potential well of the black hole. This increases the temperature of the fluid there, raising the sound speed and with it, the ISCO crossing velocity. With a larger radial velocity at the ISCO, it will take longer for the fluid to relax towards the pure geodesic solution. This trend, combined with a physically smaller plunging region, weakens the validity of the offset geodesic model as the prograde spin is increased. Despite this, since the plunging region itself is physically smaller in this limit, it is of less importance to the overall accretion scenario.
\par
In the top-right panel of Fig.\,\ref{fig:uprofiles} we have also plotted the density-weighted averaged profiles of the azimuthal component $U^\phi$. Here we must emphasise that our analyses are done in horizon-penetrating Spherical Kerr-Schild (SKS) coordinates, \emph{not} the commonly used \cite{boyerMaximalAnalyticExtension1967} coordinates. Whilst the two coordinate systems are identical for radial (contravariant) components, the azimuthal components differ:
\begin{equation*}
    U_{\mathrm{SKS}}^{\phi} = U_{\mathrm{BL}}^{\phi} + \frac{a U^r}{r^2-2r+a^2}.
\end{equation*}
We will henceforth drop all coordinate label subscripts and adopt exclusively SKS coordinates. In these coordinates, the geodesic solution for $U^\phi$ corresponding to the same energy and angular momentum as an ISCO orbit has the form (see Appendix \ref{sec:AppA}):
\begin{equation}
    U^\phi = \frac{\Delta J_I^2+2 a A J_I - a^2 B}{r^2\left(a A + \Delta J_I + a\sqrt{B \Delta + A^2}\right)},
    \label{eq:uphigeo}
\end{equation}
where
\begin{align*}
    A = 2 r \gamma_I - aJ_I, \\
    B = (\gamma_I^2 - 1)r^2 + 2 \gamma_I^2 r - J_I^2, \\
    \Delta = r^2 - 2r +a^2, \\
    \gamma_I = \sqrt{1-\frac{2}{3r_I}}, \\
    J_I = 2 \sqrt{3} \left(1-\frac{2a}{3\sqrt{r_I}}\right).
\end{align*}
\par
Unlike $U^r$, we cannot construct a single spin-independent curve from $U^\phi$. Instead, we repeatedly plot Eq.\,\ref{eq:uphigeo} with dashed lines in the top-right panel of Fig.\,\ref{fig:uprofiles}, with the appropriate black hole spin for each simulation. In all cases, we observe that the simulated profile falls slightly below the geodesic solution, clearly demonstrating the non-zero shear stress in the plunging region. The largest departures occur for the rapidly spinning prograde black holes, with the low-spin and retrograde black holes tracking the geodesic solution quite closely. This indicates that the magnetic ISCO stress is an increasing function of prograde spin. Nonetheless, as we argued in \citetalias{rulePlungingRegionThin2025a}, for all but the most extremal spinning black holes these stresses are (dynamically) small, and the dynamics of the plunging flow closely resembles geodesic inflow. Therefore, for our weakly magnetised physically thin disc simulations, we infer that gravity remains the dominant forcing term in the relativistic Euler equation. Magnetic fields and pressure gradients are dynamically sub-dominant in the plunging region across the black hole spin range.
\subsection{Testing the MB23 framework across the black hole spin range}
\label{sec:MB23Test}
\begin{figure}
    \centering
    \includegraphics[width=0.9\linewidth]{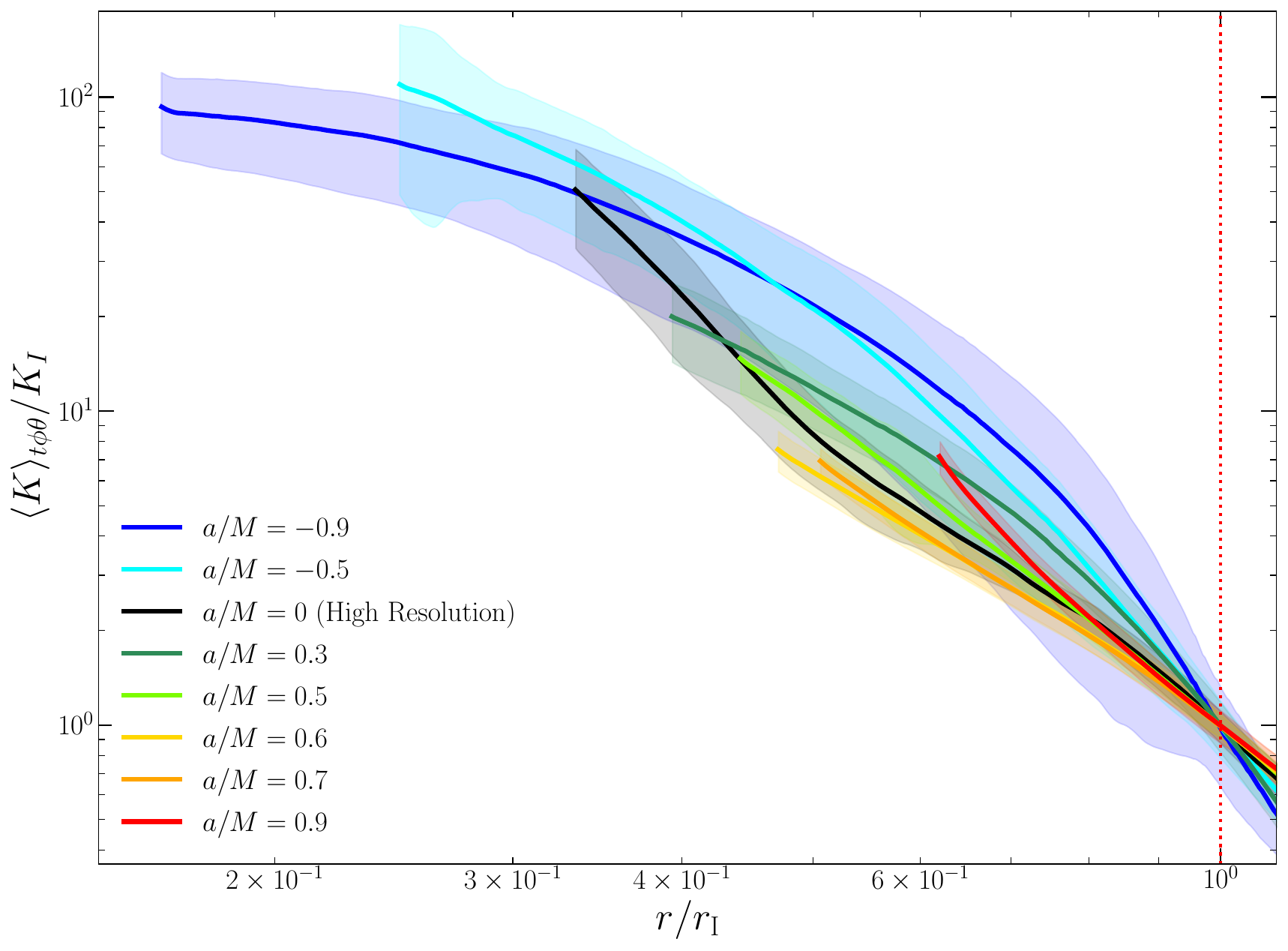}
    \caption{
    The plunging region radial $K=P\rho^{-\gamma_{\rm ad}}$ profile, normalised to the value at the ISCO. Solid lines show the normalised vertically, azimuthally and temporally averaged profiles, $\langle K \rangle_{t \phi \theta}/K_I$. The shaded regions represent a $\pm1\sigma$ standard deviation measured by the temporal variance of the spatially averaged quantities during the averaging window (see \citetalias{rulePlungingRegionThin2025a}).
    }
    \label{fig:kprofiles}
\end{figure}

Given that the fluid dynamics may be approximated by geodesic inflow, we shall next explicitly test the \citetalias{mummeryAccretionInnermostStable2023} thermodynamic framework against the set of simulations across the spin range. A key finding of \citetalias{rulePlungingRegionThin2025a} was that the plunging flow is not adiabatic, plausibly (though not definitively) due to numerical magnetic dissipation at a current sheet that forms in the mid-plane of the flow.  We found that the quantity $K = P\rho^{-\gamma_{\rm ad}}$, which is related to the entropy via $S = k_B \ln(K)/(\gamma_{\rm ad}-1)$, may be described by an ad-hoc radial power law $K = K_I (r/r_I)^{-m}$. This enables us to account for this non-adiabatic magnetic heating in the analytic thermodynamic solutions. Fig.\,\ref{fig:kprofiles} shows the simulated radial profiles for $K$. For the simulations with a prograde black hole spin, the power-law model appears to be an adequate model. Whilst both retrograde simulations (and potentially $a/M=0.3$) deviate noticeably from our ad-hoc power-law prescription, we choose to keep it for now. Indeed, it is a much better description of the simulated profiles than a constant-$K$ model would be. While it would be more desirable to use a physically motivated model for $K$, a detailed treatment of this complex physical process lies outside the scope of this work. As a simple placeholder, the power-law model does surprisingly well.
\par
\begin{figure*}
    \centering
    \includegraphics[width=0.42\linewidth]{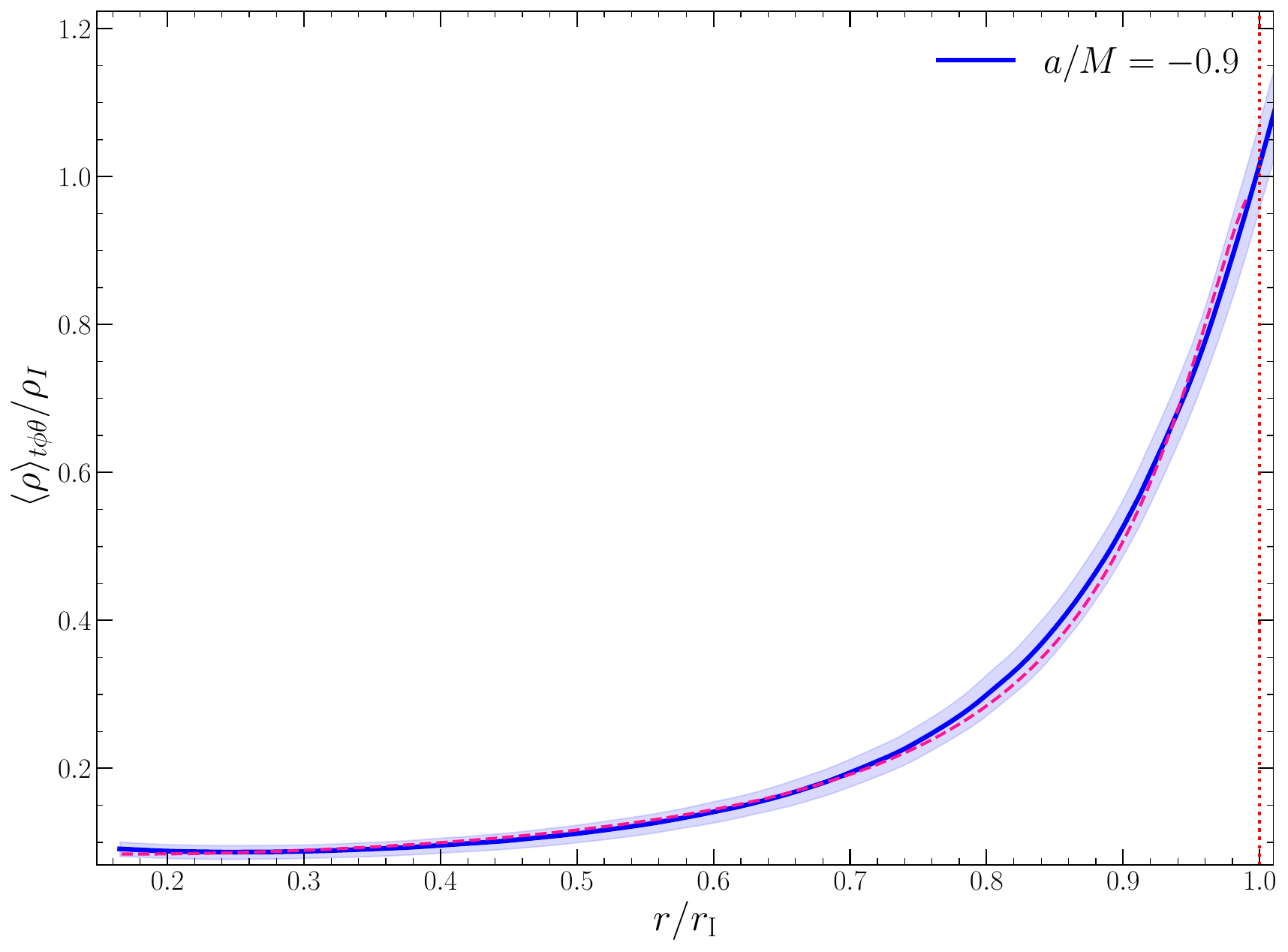}
    \includegraphics[width=0.42\linewidth]{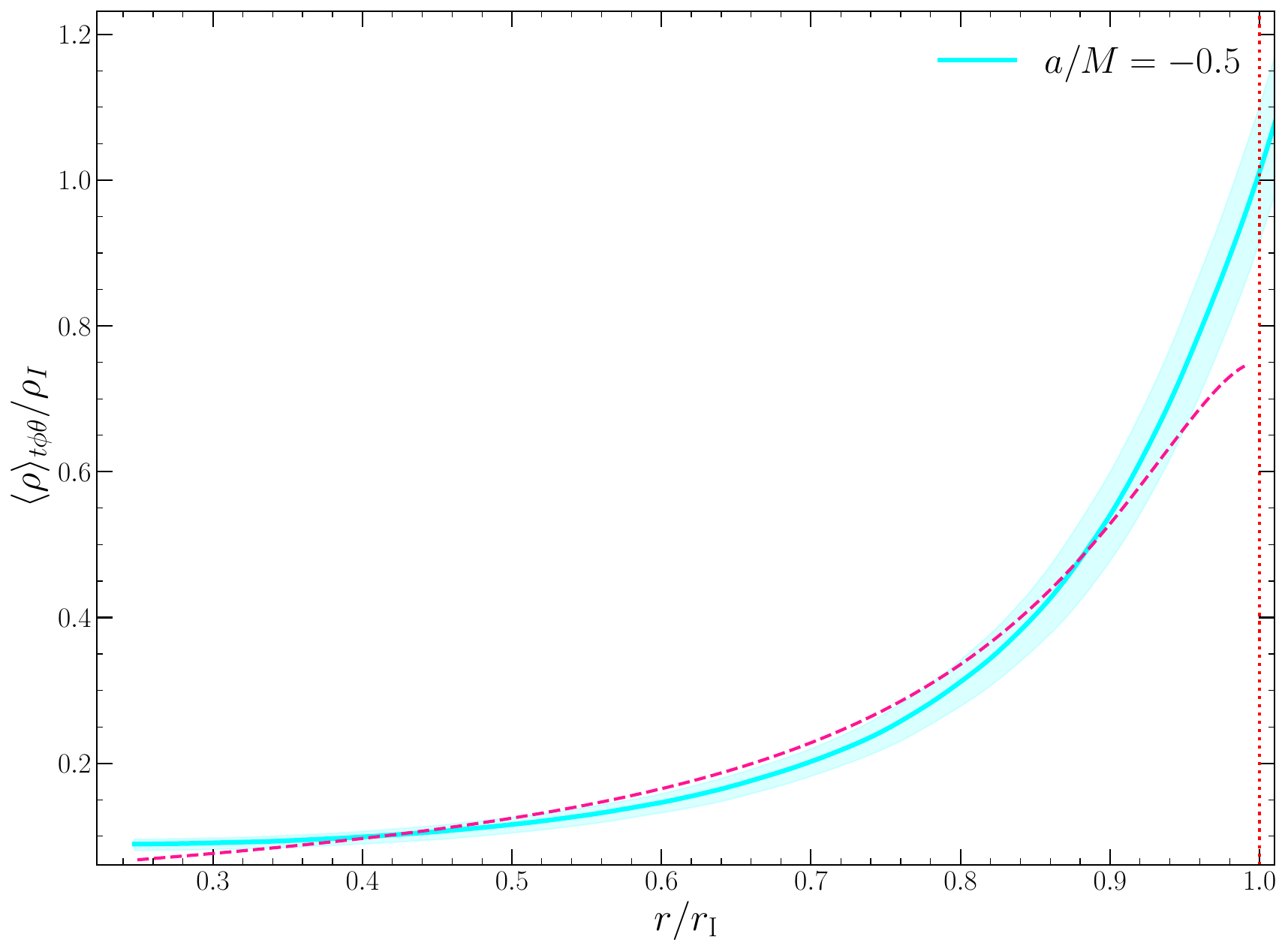}
    \includegraphics[width=0.42\linewidth]{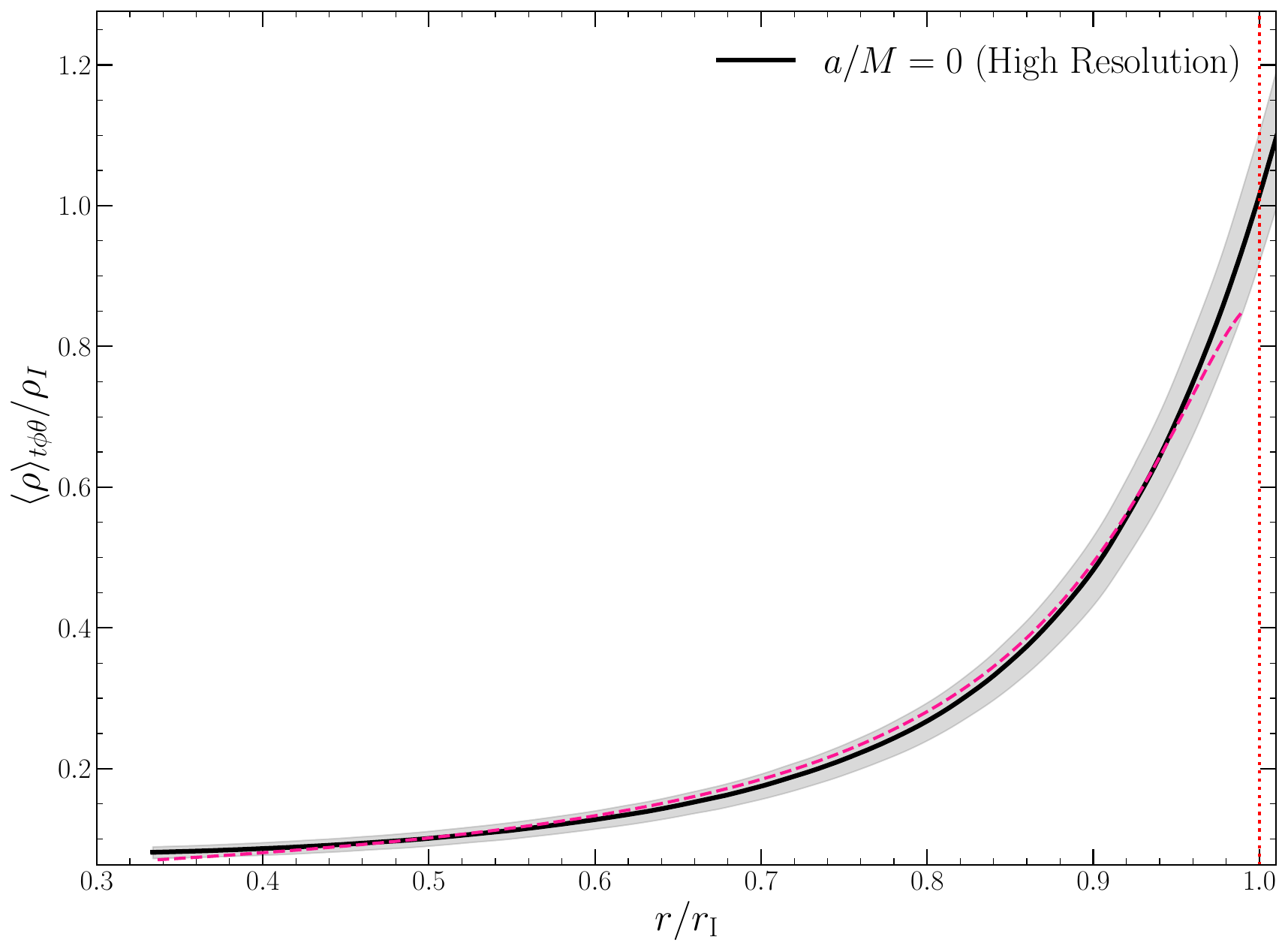}
    \includegraphics[width=0.42\linewidth]{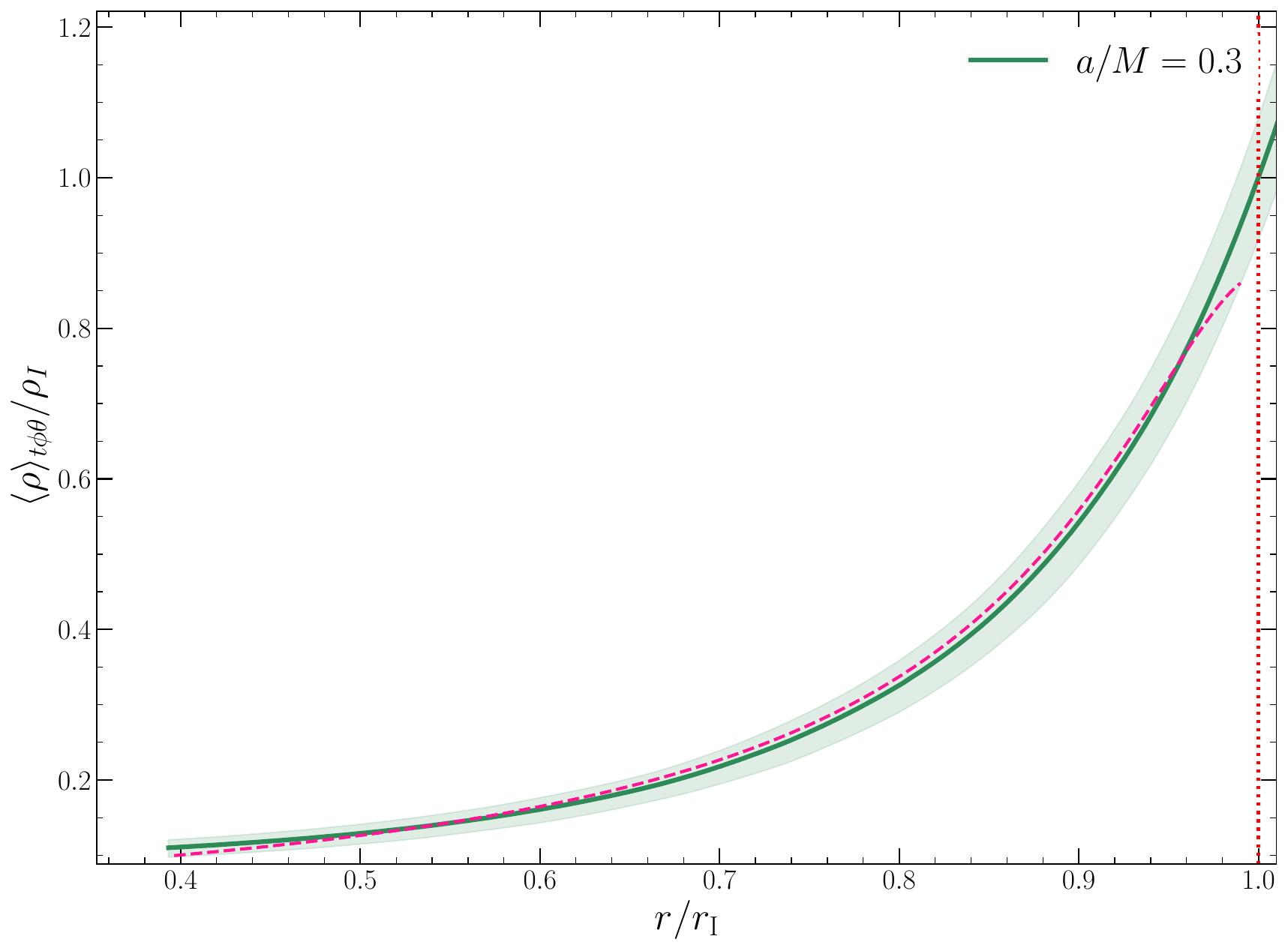}
    \includegraphics[width=0.42\linewidth]{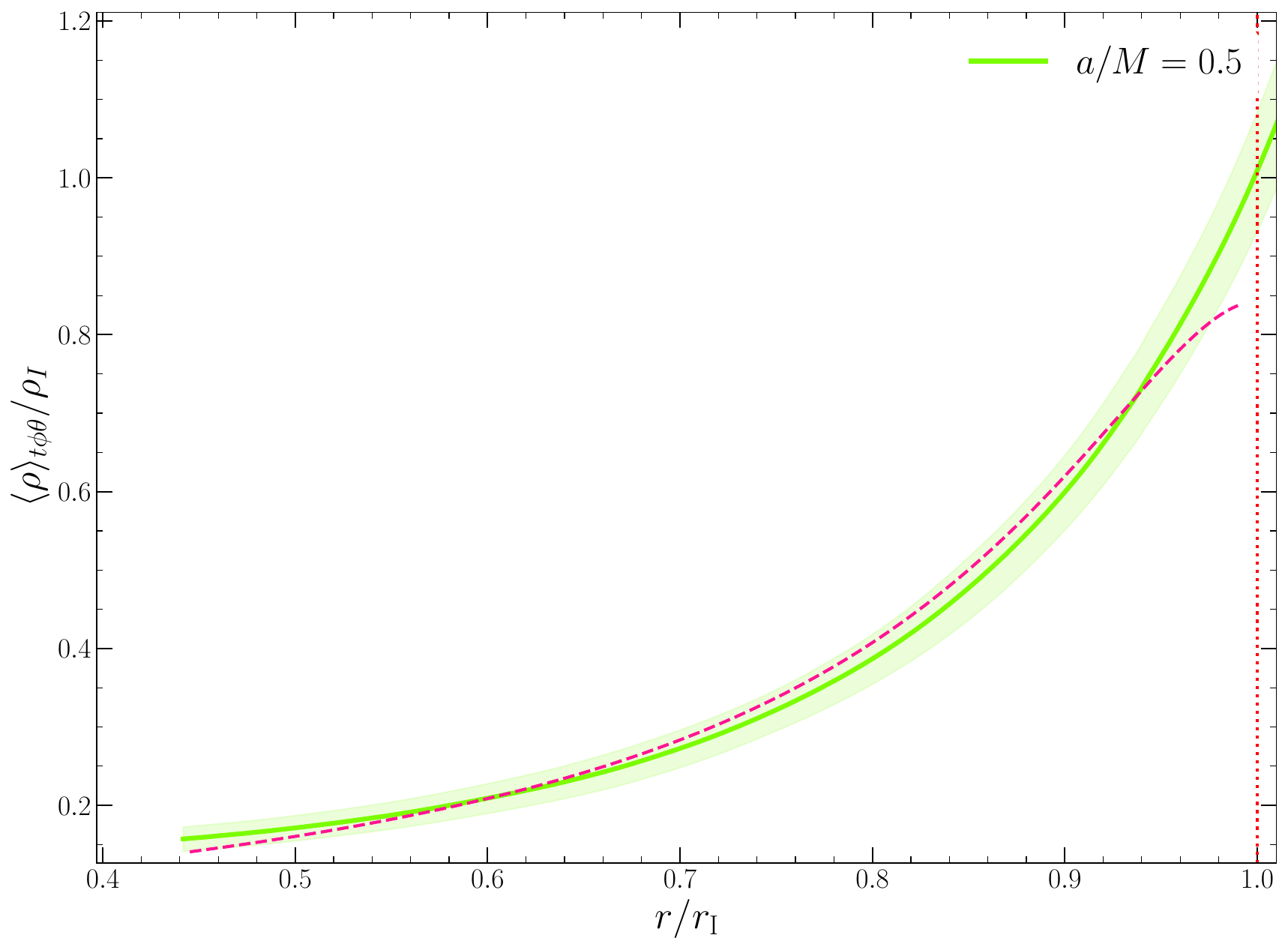}
    \includegraphics[width=0.42\linewidth]{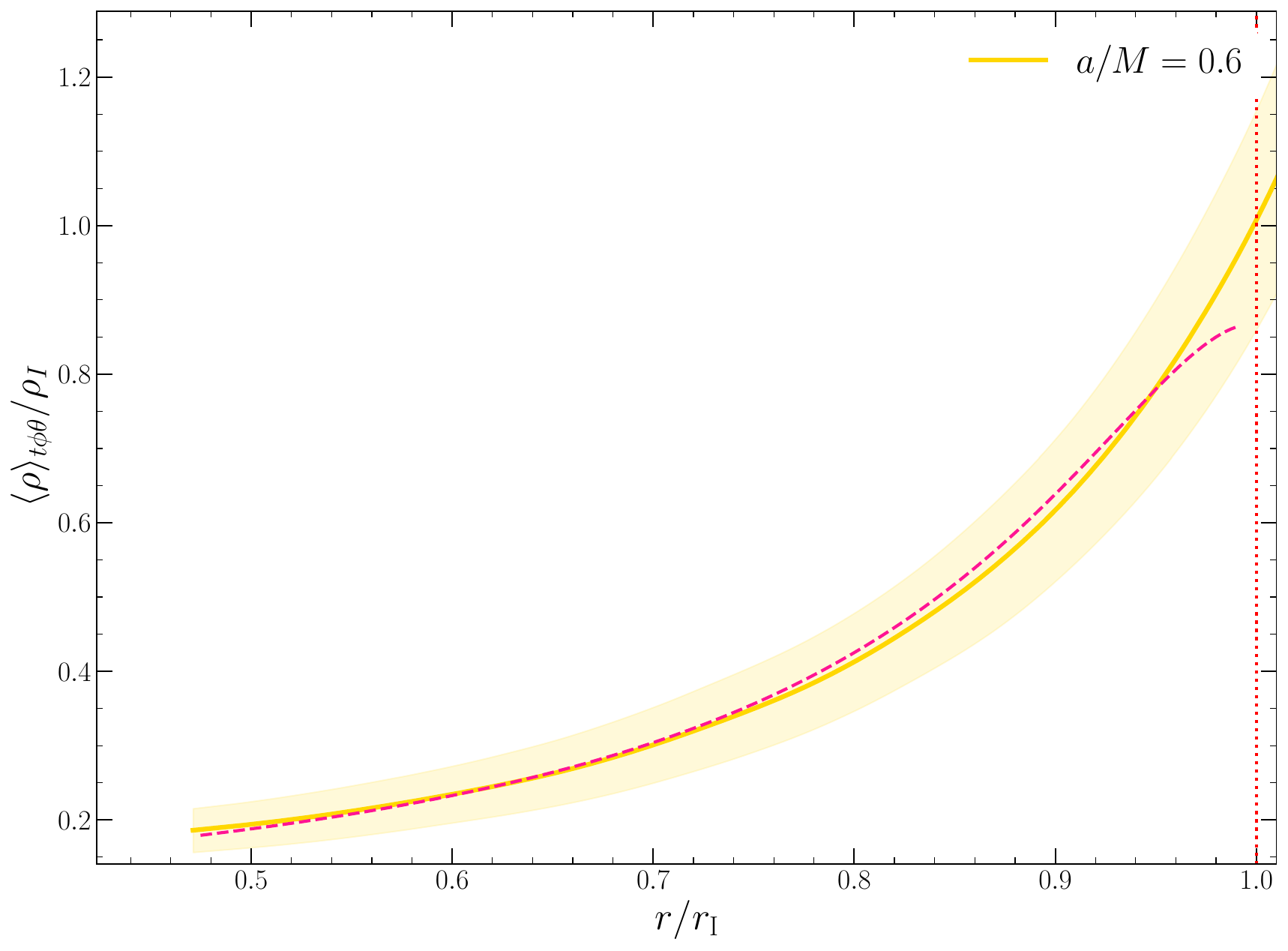}
    \includegraphics[width=0.42\linewidth]{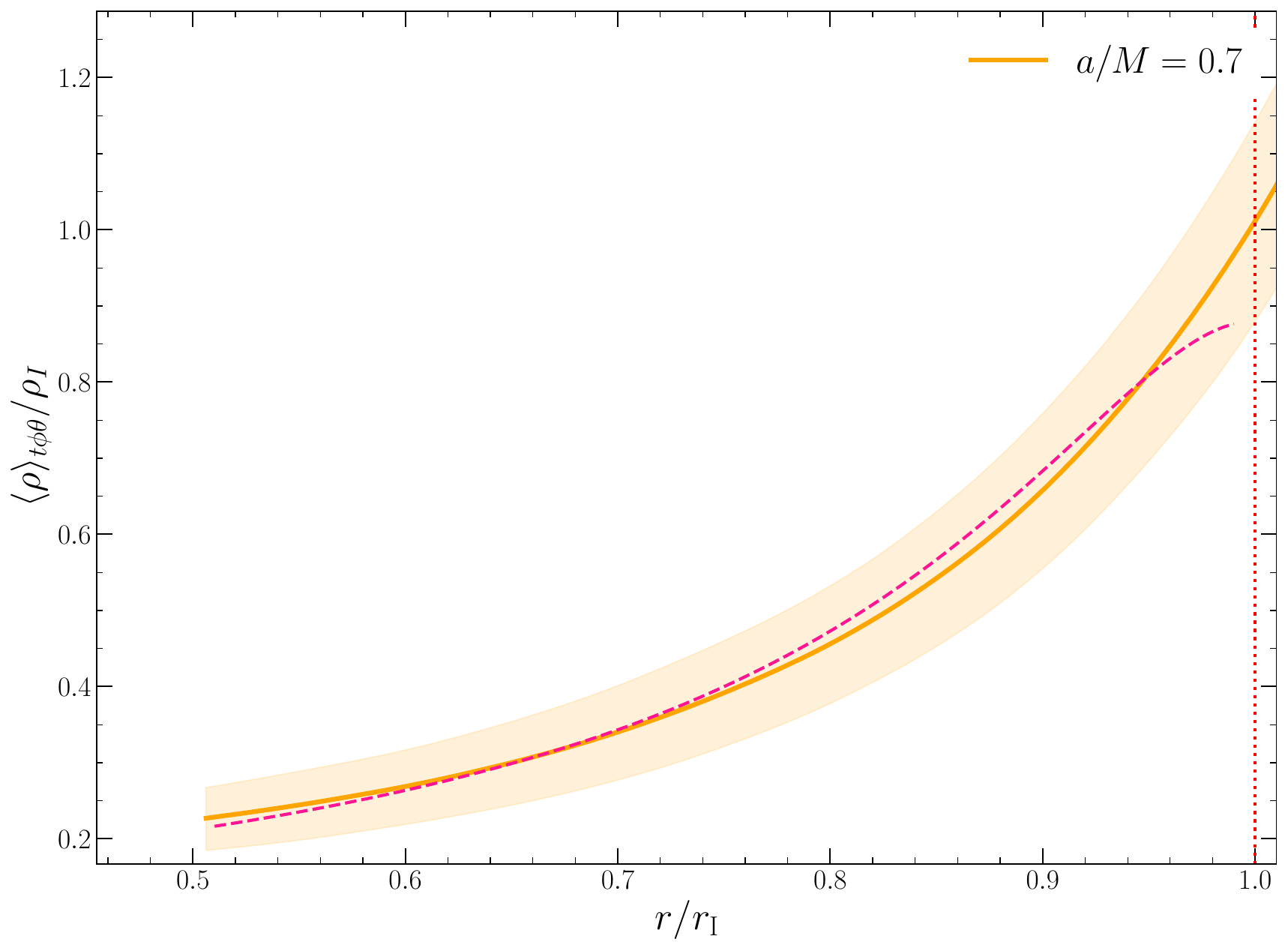}
    \includegraphics[width=0.42\linewidth]{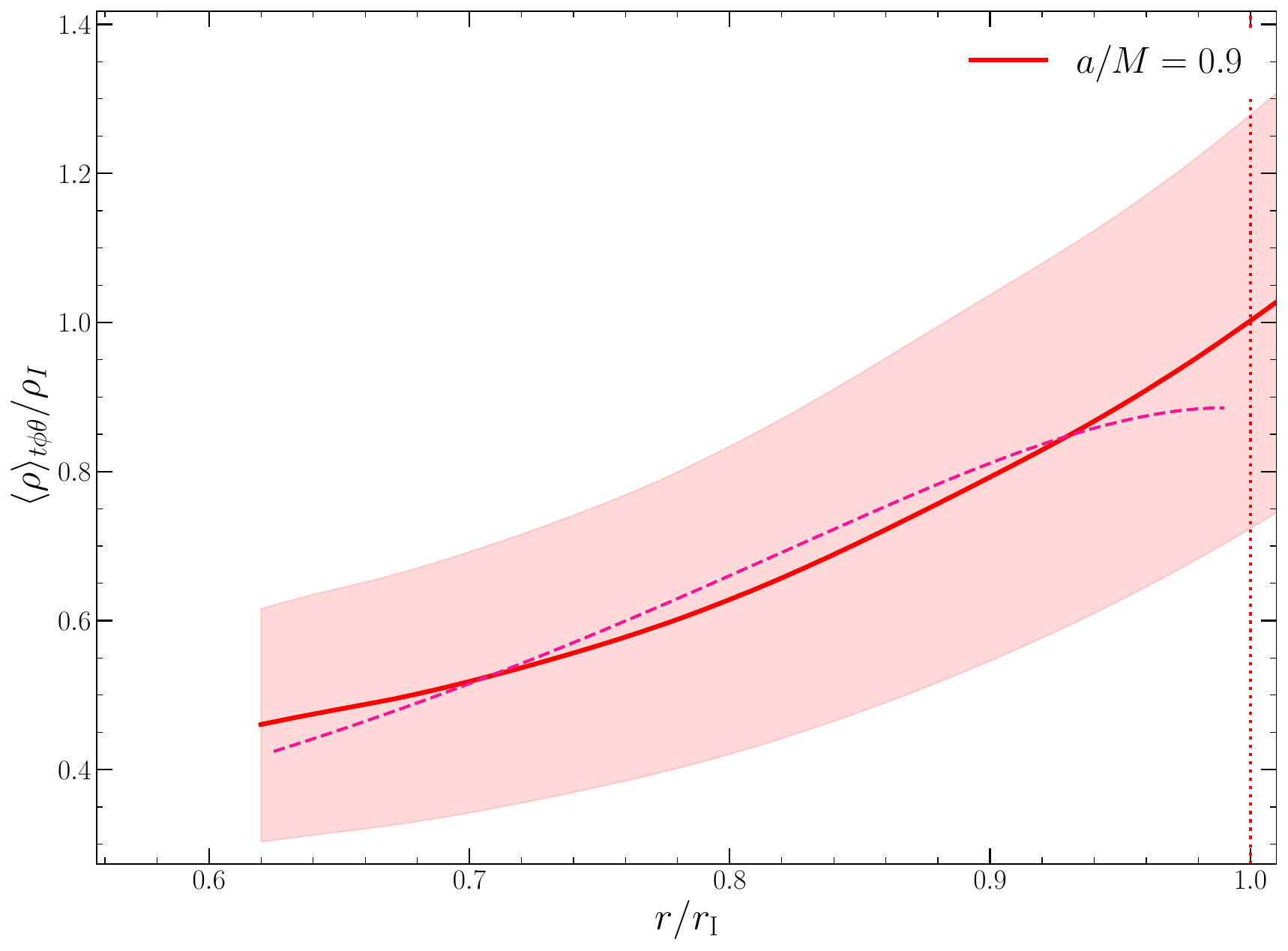}
    \caption{
    The plunging region radial density profile of each simulation, normalised to the value at the ISCO. Solid lines indicate the normalised vertically, azimuthally and temporally averaged density profiles, $\langle \rho \rangle_{t \phi \theta}/\rho_I$. The shaded regions represent a $\pm1 \sigma$ temporal standard deviation (see \citetalias{rulePlungingRegionThin2025a}). The dashed lines are fitted \citetalias{mummeryAccretionInnermostStable2023} density models for each simulated profile. It is important to note that we model $K$ with a different fitted power-law in each case.
    }
    \label{fig:rhomodels}  
\end{figure*}

Fig.\,\ref{fig:rhomodels} shows the simulated radial density profile of the plunging region for each simulation, normalised to the value at the ISCO. For each simulation, we first find a best-fit power law model for $K$ using the simulated $K$ profiles in Fig.\,\ref{fig:kprofiles} (by minimising the error-weighted squared distance between the simulated data and the model). We then hold fixed that $K$ model and use it to find a best-fit density model using the following formula from \citetalias{mummeryAccretionInnermostStable2023} (where $\epsilon \equiv (U_I/c)\sqrt{3r_I/2r_g}$; both $\epsilon$ and the overall normalisation are fitted parameters),
\begin{equation}
    \frac{\rho}{\rho_I} = \left(\frac{r_I}{r}\right)^\frac{6}{\gamma_{\rm ad}+1}\left(\frac{K}{K_I}\right)^{\frac{-1}{1+\gamma_{\rm ad}}}\left[\epsilon^{-1} \left(\frac{r_I}{r}-1\right)^\frac{3}{2}+1\right]^{-\left(\frac{2}{\gamma_{\rm ad}+1}\right)}.
    \label{eq:densitymodel}
\end{equation}
These models are over-plotted in Fig.\,\ref{fig:rhomodels} with dashed lines. As noted above, our power-law model for $K$ is heuristic and appears to be weakest for the retrograde spin simulations. Nonetheless, since $\rho$ is only rather weakly dependent on $K$ (for $\gamma_{\rm ad}=13/9$, $\rho \sim K^{-9/22}$), we are able to find good fits for the density. As the black hole spin is increased in the prograde direction, our $K$ model becomes more accurate, further improving the quality of the density model, until the increasing ISCO stress undermines the assumption that the fluid follows geodesic trajectories.
\par
We find the best agreement for the low-spin (in absolute terms) simulations, where the dynamics most closely align with geodesic inflow and our power-law model for $K$ is most accurate. For these simulations, this agreement suggests that the remaining \citetalias{mummeryAccretionInnermostStable2023} assumptions, such as mass conservation and vertical hydrostatic equilibrium, provide an effective description of the plunging flow. For the rapidly spinning prograde simulations, the disc is considerably hotter and the plunging region is physically much smaller (at a fixed numerical resolution), so the averaged density profiles are more variable. For these simulations, the best-fit density models appear to be accurate to within the error bounds. However, for $a/M=0.9$, this may be a closer reflection of the variability rather than the accuracy of the model, especially given the departures from pure geodesic inflow that were discussed previously.
\par
Whilst the \citetalias{mummeryAccretionInnermostStable2023} density profiles in Fig.\,\ref{fig:rhomodels} are generally rather accurate for the majority of the plunge, they do not describe behaviour close to the ISCO very well.  Here, the model is fundamentally limited by the constant offset approximation that \citetalias{mummeryAccretionInnermostStable2023} employ to describe $U^r$ at the ISCO, where the geodesic term vanishes. Indeed, it may be described as a zeroth-order expansion (in $r$, about $r_I$) of an unknown transition profile from the main disc's inward radial drift to a geodesic inflow solution. As a mathematical construction, a non-zero radial velocity at the ISCO is of course essential to avoid an infinite surface density at the ISCO while retaining a finite mass accretion rate. We do not expect the constant offset model to be a particularly good physical description of the complex underlying physics of the transition region, which is why the \citetalias{mummeryAccretionInnermostStable2023} density profiles all break down there. In fact, we assume that as the geodesic solution rapidly grows, it will quickly dominate over the transition profile, greatly suppressing our dependence on it for the majority of the plunging region. Nonetheless, it is clearly a promising avenue for further research to develop better models for the transition profile, so that we may improve our modelling in the neighbourhood of the ISCO itself.
\par
Overall, we find good agreement between the best-fit density model (i.e. Equation\,\ref{eq:densitymodel}) and the simulated profiles across the spin range. The \citetalias{mummeryAccretionInnermostStable2023} framework appears to be useful for analysing the plunging flow in observational contexts where the black hole spin may be unknown, so long as proper care is taken for the interpretation of results close to the extreme spins. 
\subsection{The ISCO Stress as a function of black hole spin}
\label{sec:ISCOStress}
\begin{figure*}
    \centering
    \includegraphics[width=0.49\linewidth]{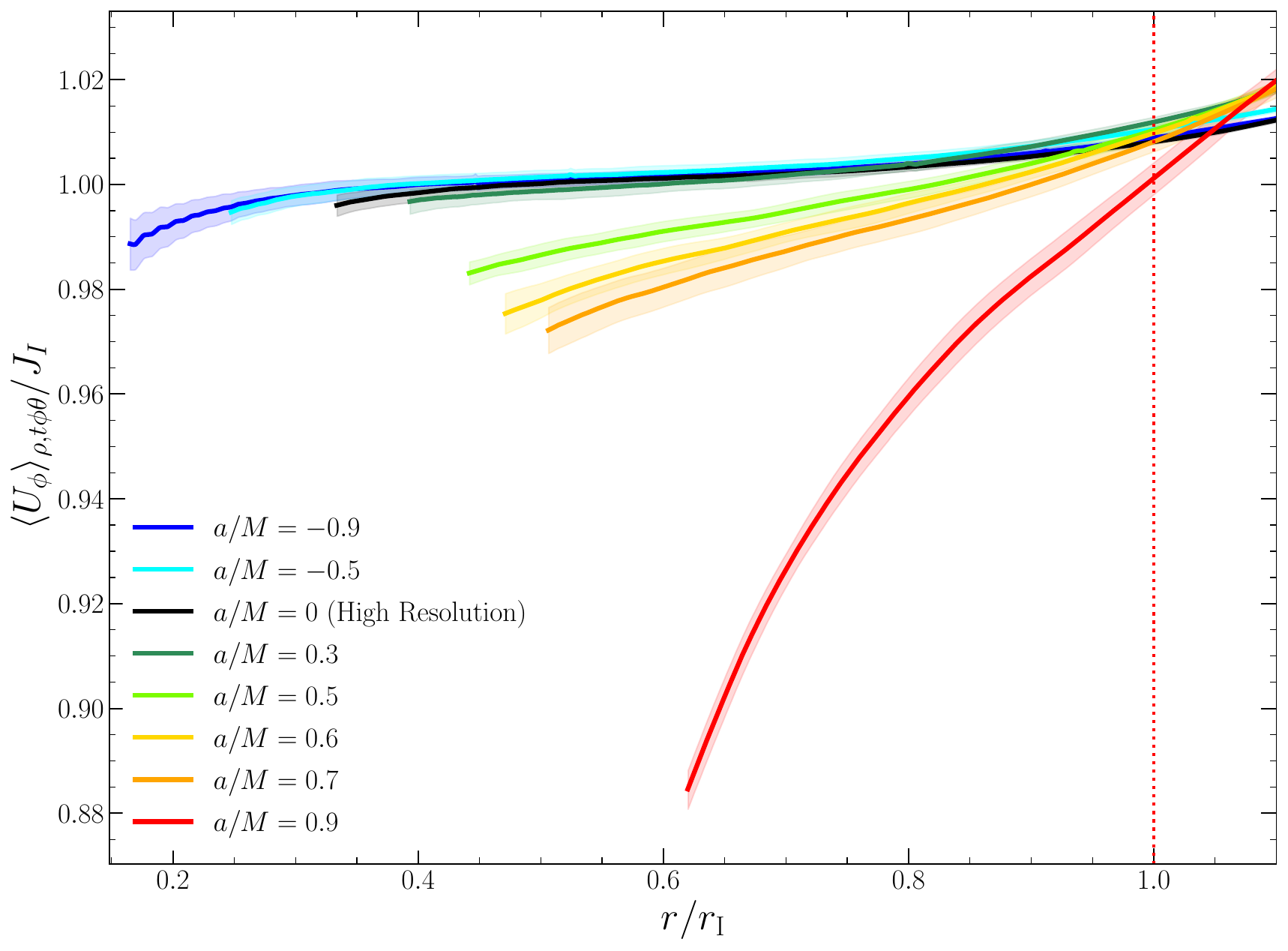}
    \includegraphics[width=0.49\linewidth]{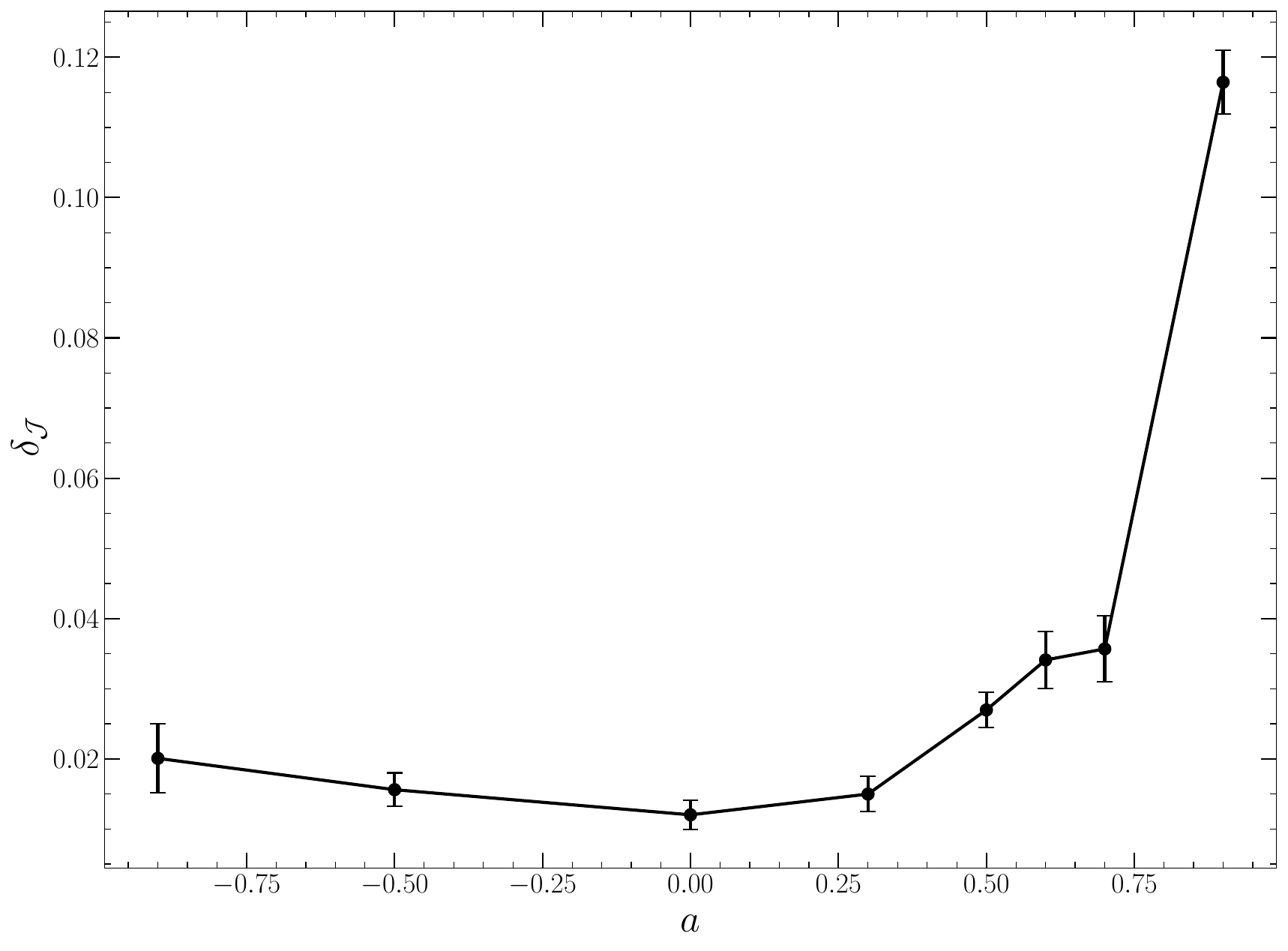}
    \caption{The specific angular momentum $U_\phi$, normalised by the angular momentum of a circular orbit at the ISCO $J_I$ (left) and the fractional change in the specific angular momentum as a function of spin, $\delta_\mathcal{J}$ (right). The solid lines on the left are the density-weighted, temporally, azimuthally and vertically averaged quantities for each simulation. The shaded intervals represent a $\pm 1\sigma$ standard deviation.}
    \label{fig:ISCOStress}   
\end{figure*}

The left-hand side of Fig.\,\ref{fig:ISCOStress} shows the density weighted average specific angular momentum ($U_\phi$) as a function of radius for each simulation. On the right-hand side, we plot the fractional drop in this quantity from the ISCO to the event horizon as a function of spin. Immediately, we notice that this drop rises sharply as the prograde spin is increased. There is also a very modest rise in the retrograde direction from $a/M=0.0$, which is the minimum. These observations may be explained by two competing effects. First of all, our retrograde simulation, with $a/M=-0.9$, has the physically largest plunging region. As the spin is increased in the prograde direction, the extent of the plunging region shrinks. There is less time for the shear stresses (i.e. the $T^r_\phi$ stress tensor component) to act, and so the accumulated drop in angular momentum reduces.
\par
\begin{figure}
    \centering
    \includegraphics[width=0.95\linewidth]{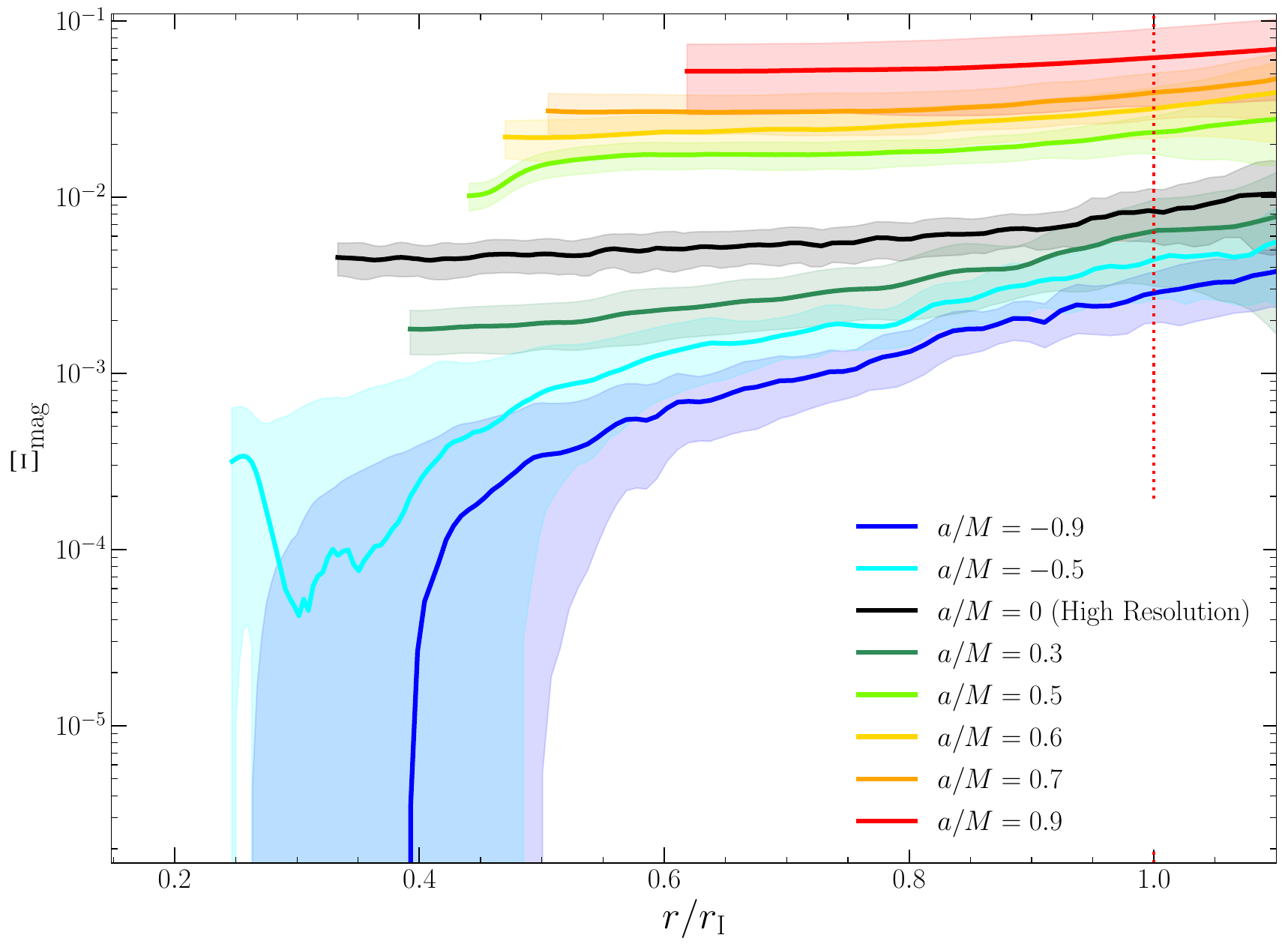}
    \caption{Time averaged radial profiles of the shell integrated radial flux of angular momentum ($\dot{J}_\mathrm{mag}$) of magnetic origin normalised by the mass accretion rate ($\dot{M}$) for each simulation. The shaded region indicates a $\pm1\sigma$ standard deviation.}
    \label{fig:GlobalFluxes}   
\end{figure}

The competing effect is that there is also a rapid increase in the magnetic stress as a function of increasing spin in the prograde direction. Despite acting over a shorter interval of time, as the prograde spin increases, there is a sharp rise in the accumulated drop in angular momentum. To study this trend, it is helpful to track the flux of angular momentum of magnetic origin through radial shells for each simulation. The relevant time-averaged shell-integrated flux is the integral of $-b^r b_\phi$, the magnetic part of $T^r_\phi$, the relevant component of the stress-energy tensor. We integrate over surfaces of constant $r$ centred at the origin:
\begin{equation}
    \dot{J}_\mathrm{mag}(r) = \left. \frac{1}{t_f-t_i} \int_{t_i}^{t_f} \int_0^\pi \int_0 ^{2\pi} \left ( - \sqrt{\vert g \vert} b^r b_\phi  \right) \mathrm{d}\phi \mathrm{d}\theta \mathrm{d}t \right \vert_r . 
\end{equation}
In Fig.\,\ref{fig:GlobalFluxes}, we plot $\Xi_\mathrm{mag}(r) \equiv \dot{J}_\mathrm{mag}(r)/(J_I \dot{M}(r))$, which is the dimensionless flux of angular momentum through a 2-sphere of radius $r$ that we define. To normalise the raw flux, we have used $J_I$, which is the specific angular momentum ($U_\phi$) of a circular orbit at the ISCO. We also compute the time-averaged shell integrated mass accretion rate:
\begin{equation}
    \dot{M}(r) = - \left. \frac{1}{t_f-t_i} \int_{t_i}^{t_f} \int_0^\pi \int_0 ^{2\pi} \left ( \sqrt{\vert g \vert} \rho U^r \right) \mathrm{d}\phi \mathrm{d}\theta \mathrm{d}t \right \vert_r.
\end{equation}
In these expressions, we have introduced $b^\mu$, which is the magnetic field four-vector \citep[e.g.][]{gammieHARMNumericalScheme2003}. This vector is defined from the electromagnetic tensor $F^{\mu \nu}$ in the following way:
\begin{equation}
    b^\mu = \frac{1}{2}\epsilon^{\mu \nu \alpha \beta} U_\nu F_{\alpha \beta}, 
\end{equation}
where $\epsilon^{\mu \nu \alpha \beta}$ is the antisymmetric Levi-Civita tensor \footnote{Note that it therefore follows that $b^\mu U_\mu = 0$.}.
\par
\begin{figure*}
    \centering
    \includegraphics[width=0.49\linewidth]{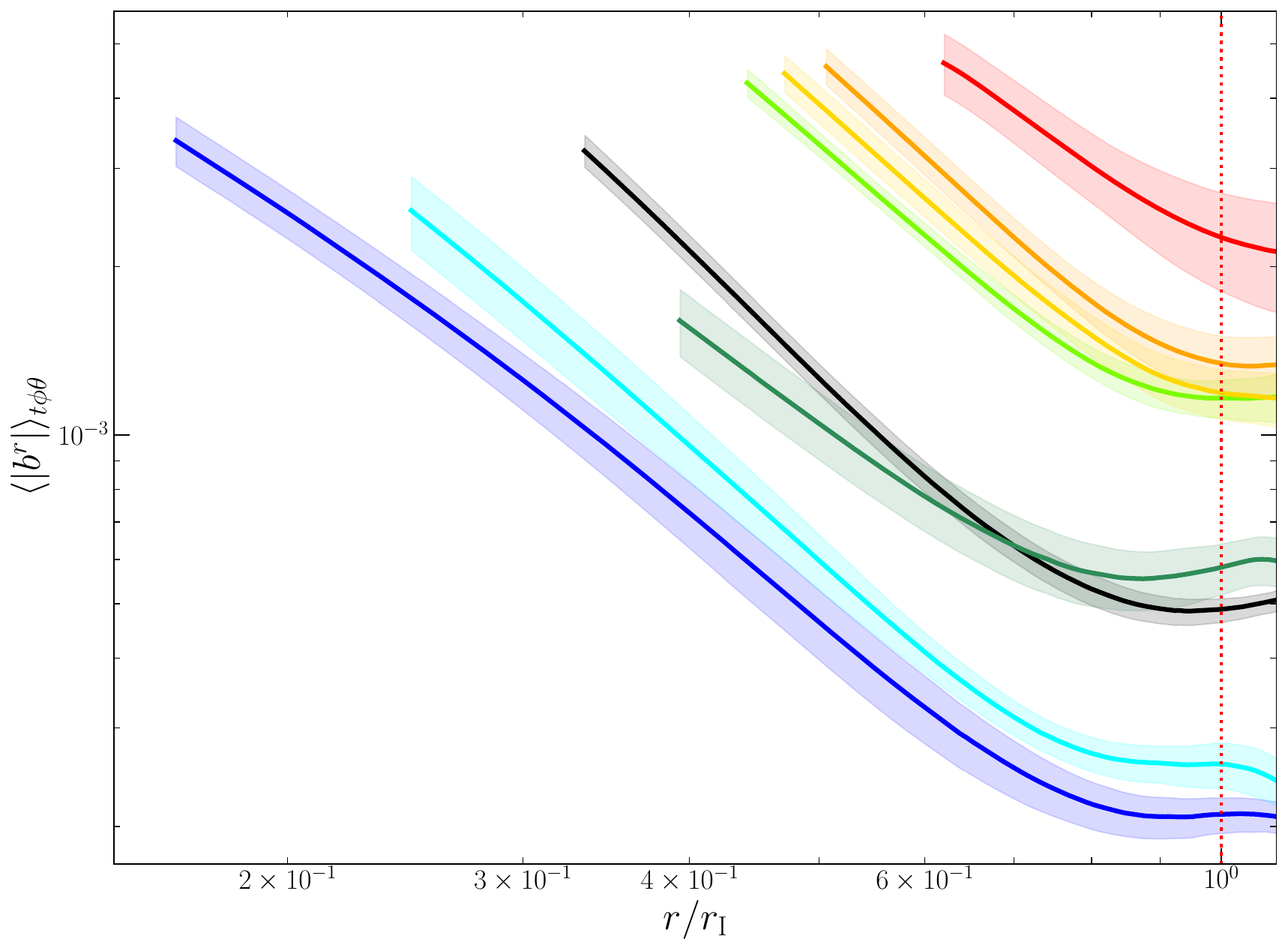}
    \includegraphics[width=0.49\linewidth]{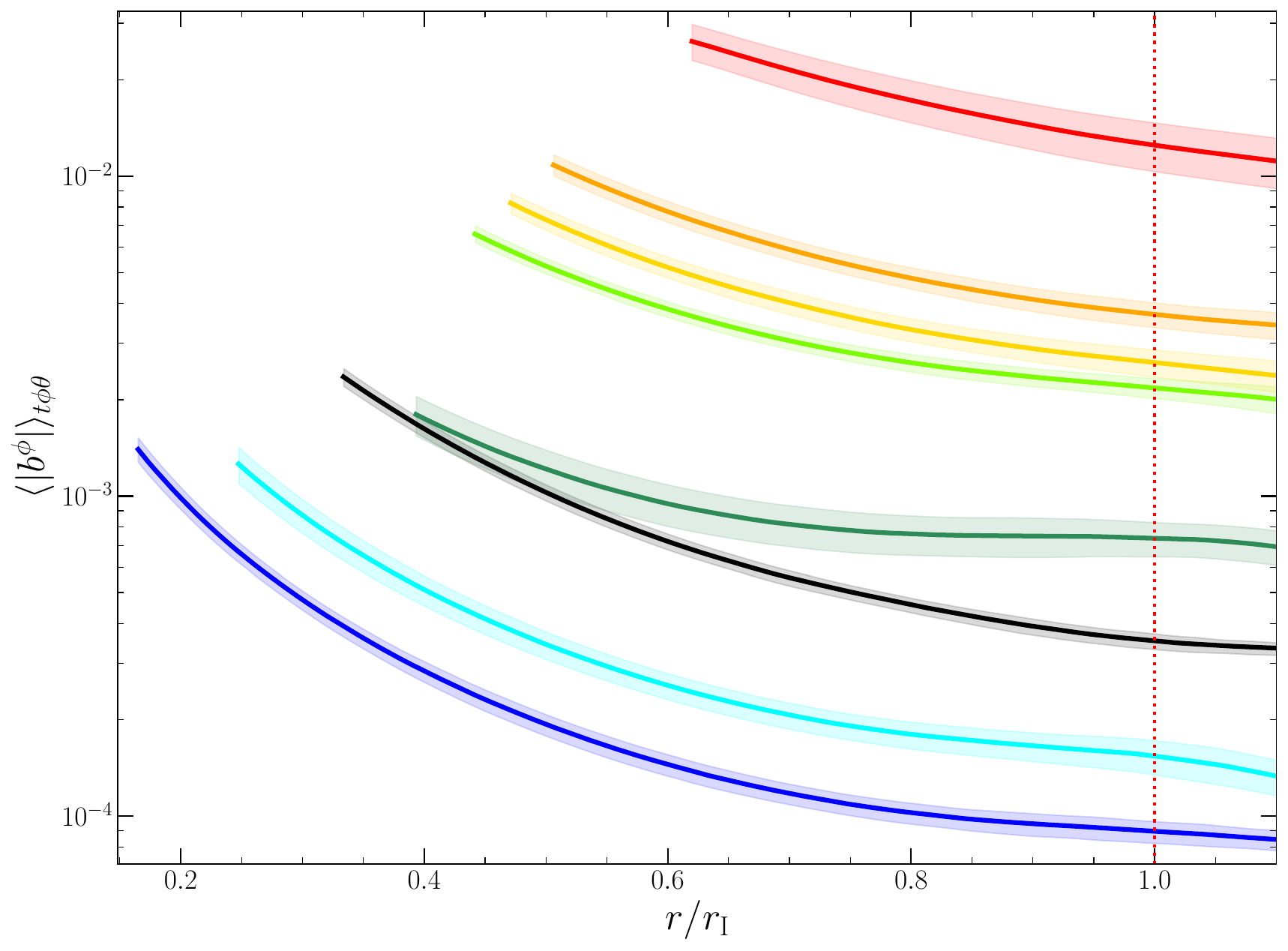}
    \includegraphics[width=0.49\linewidth]{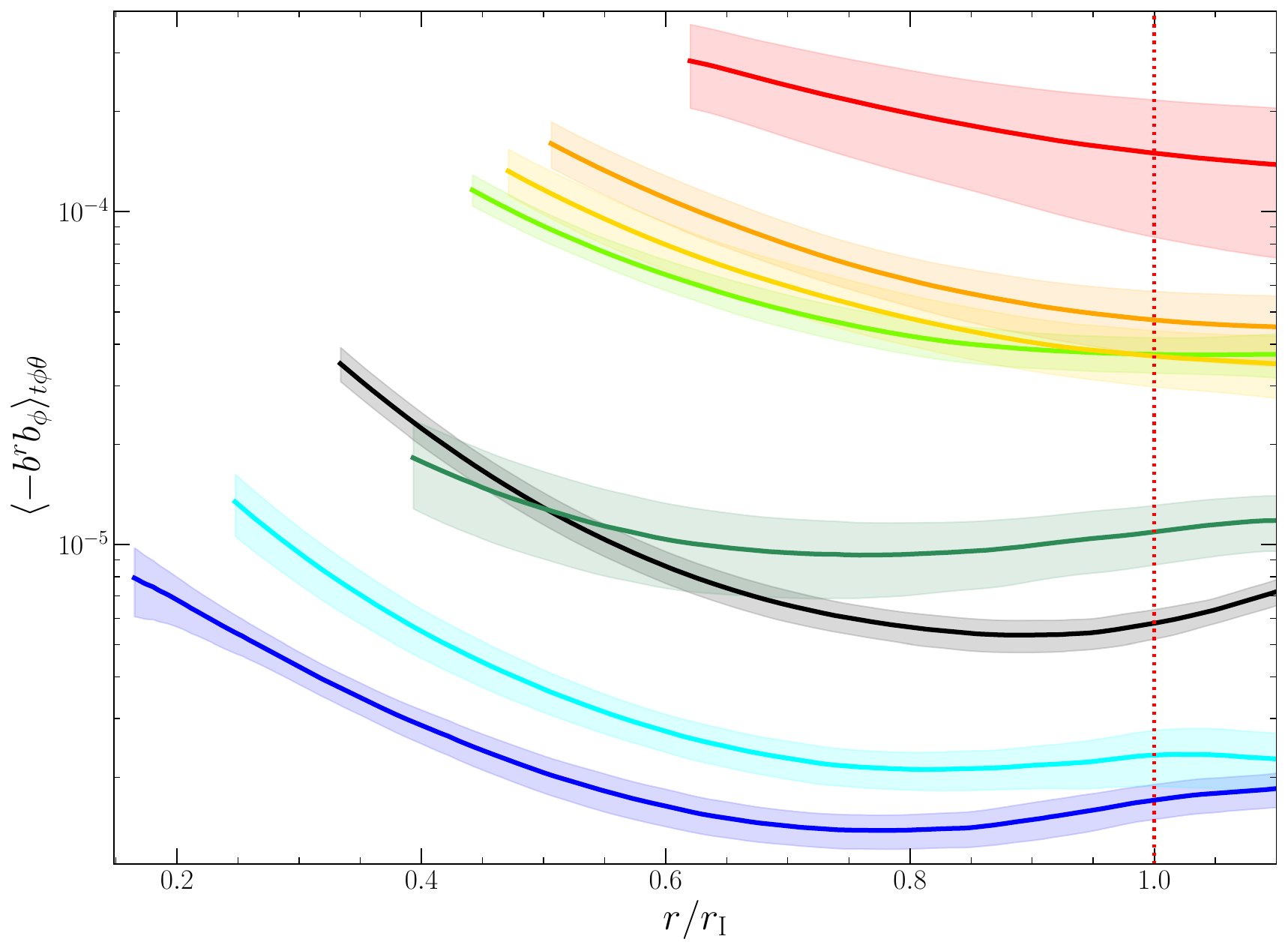}
    \includegraphics[width=0.49\linewidth]{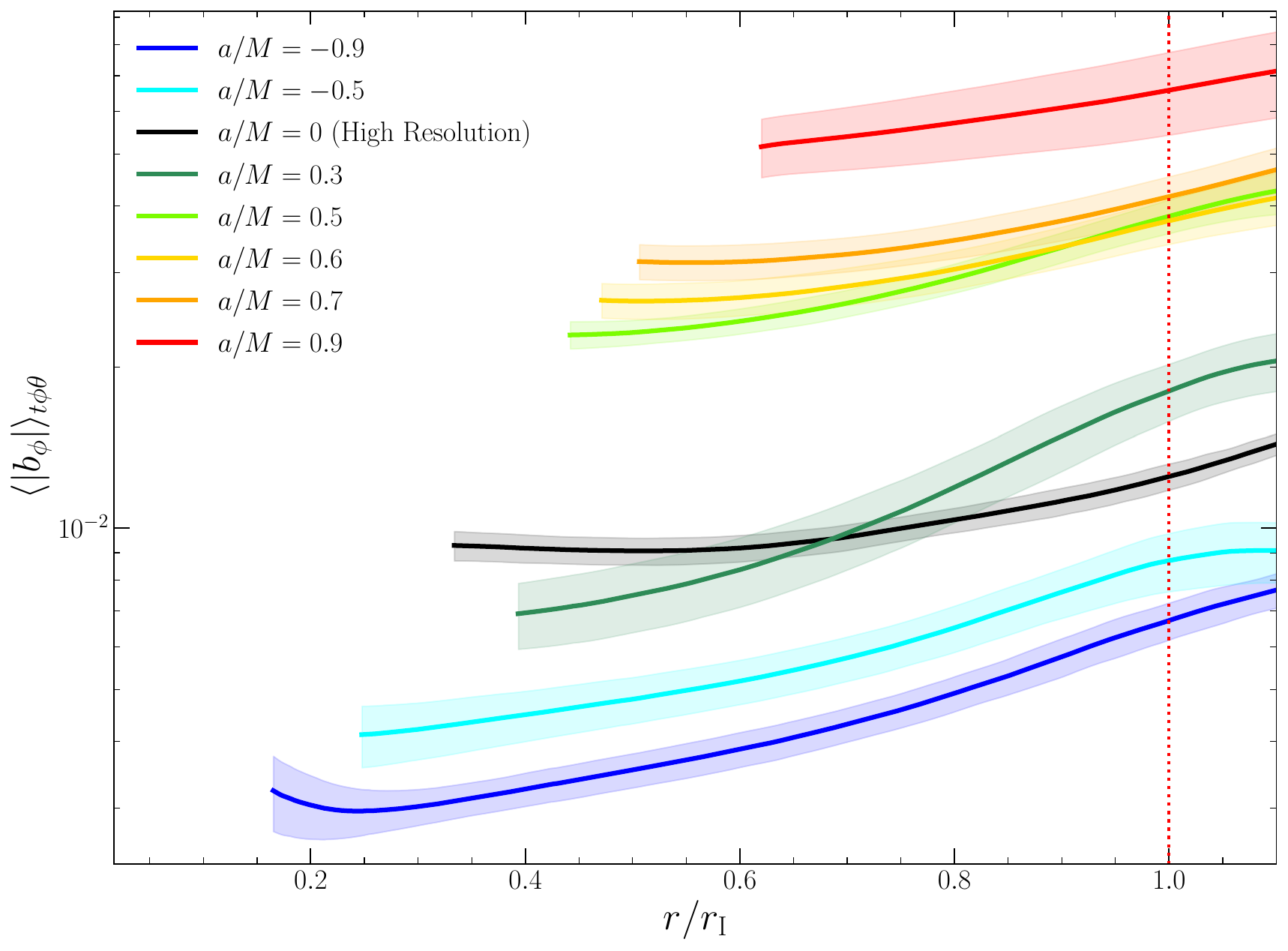}
    \caption{Clockwise starting from the top-left: the magnitude of the $r$ and $\phi$ contravariant components of the magnetic 4-vector ($\vert b^r \vert$ and $\vert b^\phi \vert$). $\vert b_\phi \vert$, the magnitude of the covariant $\phi$ component of the same vector (bottom-right). Finally, the $r -\phi$ component of the magnetic stress-energy tensor: ${T^{r}_{\phi}}_\mathrm{mag} = -b^r b_\phi$ (bottom-left). Each solid line is a vertically, azimuthally and temporally averaged radial profile for each quantity, for each simulation. The shaded regions are $\pm 1 \sigma$ standard deviations.
    }
    \label{fig:MagneticFields}   
\end{figure*}

Despite the predominantly inward advection of angular momentum with the flow, some is carried back outward by the magnetic stresses that we have repeatedly invoked in previous sections. In Fig.\,\ref{fig:GlobalFluxes}, we track these magnetic stresses explicitly. There is an (almost) systematically increasing outward (positive) flux of angular momentum as a function of increasing prograde spin \footnote{This trend is not quite absolute: we find that our high-resolution Schwarzschild simulation has a larger outward flux of angular momentum than the simulation run with $a/M=0.3$. However, in the bottom-left of Fig.\,\ref{fig:MagneticFields}, we plot the averaged stress, $\langle - b^r b_\phi \rangle$, and no longer observe the discrepant ordering. This is because we exclude the polar regions from this calculation. We therefore conclude that the discrepancy lies with inflow in the polar regions rather than the development of stresses in the mid-plane plunging flow.}. This is in accord with our previous result that the drop in specific angular momentum from the ISCO to the horizon rose sharply as a function of increasing prograde spin, despite the shrinking physical size of the plunging region.
\par
In Fig.\,\ref{fig:MagneticFields}, we explicitly track the individual components of the magnetic four-vector. It should first be noted that we must plot averages of the absolute values of the individual components. The reason for this is that, whilst we expect flux-freezing to imprint the large-scale structure of the inflowing fluid, there is still a sign ambiguity (alignment or anti-alignment with respect to the flow) that is retained on the scale of individual fluid parcels from the initially turbulent field in the main disc. If we were to average the signed field components, these aligned and anti-aligned contributions would cancel each other out.
\par
For radial magnetic angular momentum transport, the relevant magnetic stress is ${T^{r}_{\phi}}_\mathrm{mag}=-b^r b_\phi$ (see the bottom-left panel of Fig.\,\ref{fig:MagneticFields}). The covariant component $b_\phi$ may be expanded in terms of contravariant terms: $b_\phi = g_{0 \phi} b^0 + g_{r \phi} b^r + g_{\phi \phi} b^\phi$ and can be loosely thought of as the angular momentum of the field itself (in analogy with $U_\phi$ for the fluid). 
\par
Magnetic fields are amplified by extracting energy from kinematic velocity gradients in the fluid. This occurs via the stretching of field lines that connect fluid elements on adjacent, differentially flowing trajectories. Just as in the main body of the disc, there is a significant radial shear to the azimuthal flow throughout the ISCO transition region and into the plunging region (see the top-right panel of Fig.\,\ref{fig:uprofiles}). The existing poloidal field ($b^r$) is deformed by this shear,  seeding the growth of a stronger toroidal field ($b^\phi$), directly tapping the kinetic energy of the shearing flow. In addition to this, there is sharp radial acceleration after the ISCO transit. We find that this acceleration produces a vertically stratified radial flow (see \citetalias{rulePlungingRegionThin2025a}), introducing vertical shear. Analogously, this gradient may be tapped to directly amplify the radial field by stretching the existing vertical field.
\par
Why do faster prograde spinning black holes build up larger magnetic stresses? Amplification of the azimuthal field by shear is important. This can be seen by the ordering of the covariant $b_\phi$ component in the bottom-right of Fig.\,\ref{fig:MagneticFields}. To understand this trend, it is helpful to expand $b_\phi$ in terms of the contravariant components of the magnetic four-vector. Explicitly writing out the metric couplings in Spherical Kerr-Schild coordinates (setting $\theta=\pi/2$),
\begin{equation}
    b_\phi = \left(r^2+a^2+\frac{2Ma^2}{r}\right) b^\phi -\frac{2Ma}{r} b^0 - a \left(1 + \frac{2M}{r} \right) b^r .
\end{equation}
Except near the horizon and for large black hole spin, the contribution from the $b^\phi$ term will dominate. Moreover, in Fig.\,\ref{fig:MagneticFields}, we see that with the exception of the retrograde simulations, $b^\phi$ is considerably stronger than $b^r$ throughout the plunging region \footnote{$b^0$ (not shown in Fig.\,\ref{fig:MagneticFields}) is about the same size as $b^\phi$ and has qualitatively similar behaviour (see Section \ref{sec:Flux-Freezing}).}. We will therefore treat the behaviour of $b^\phi$ as the major contributor to $b_\phi$.
\par
\begin{figure*}
    \centering
    \includegraphics[width=0.33\linewidth]{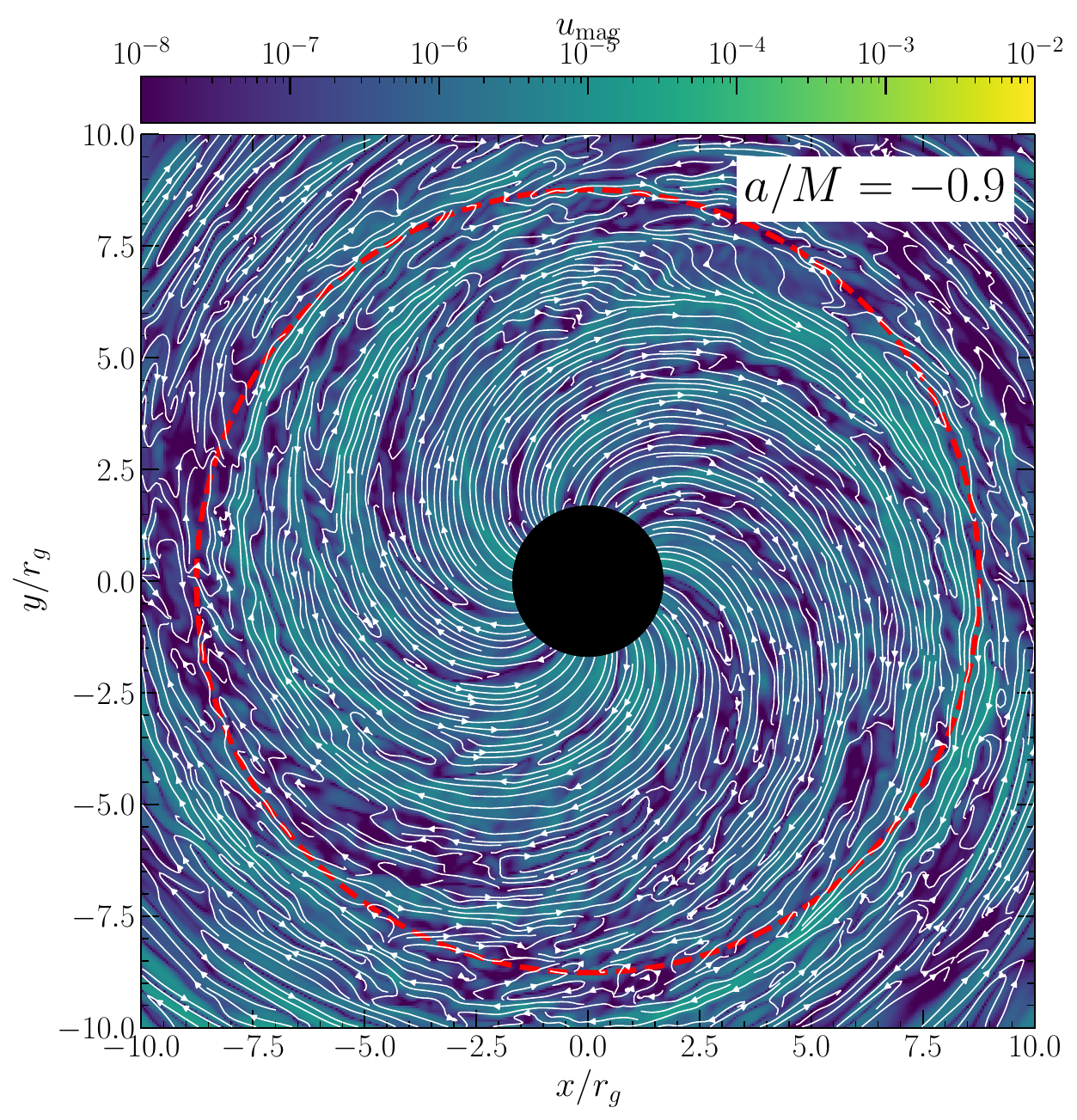}
    \includegraphics[width=0.33\linewidth]{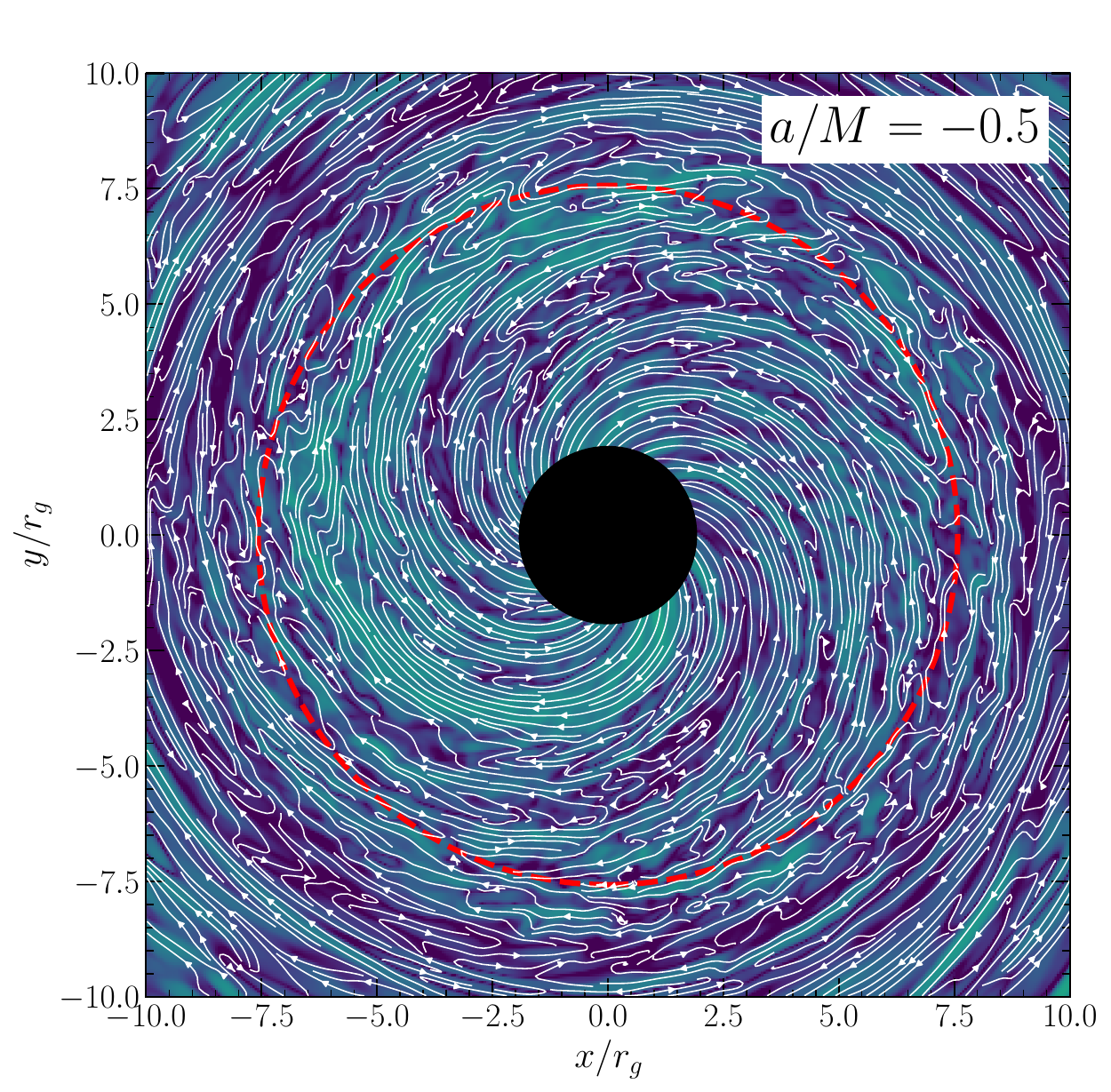}
    \includegraphics[width=0.33\linewidth]{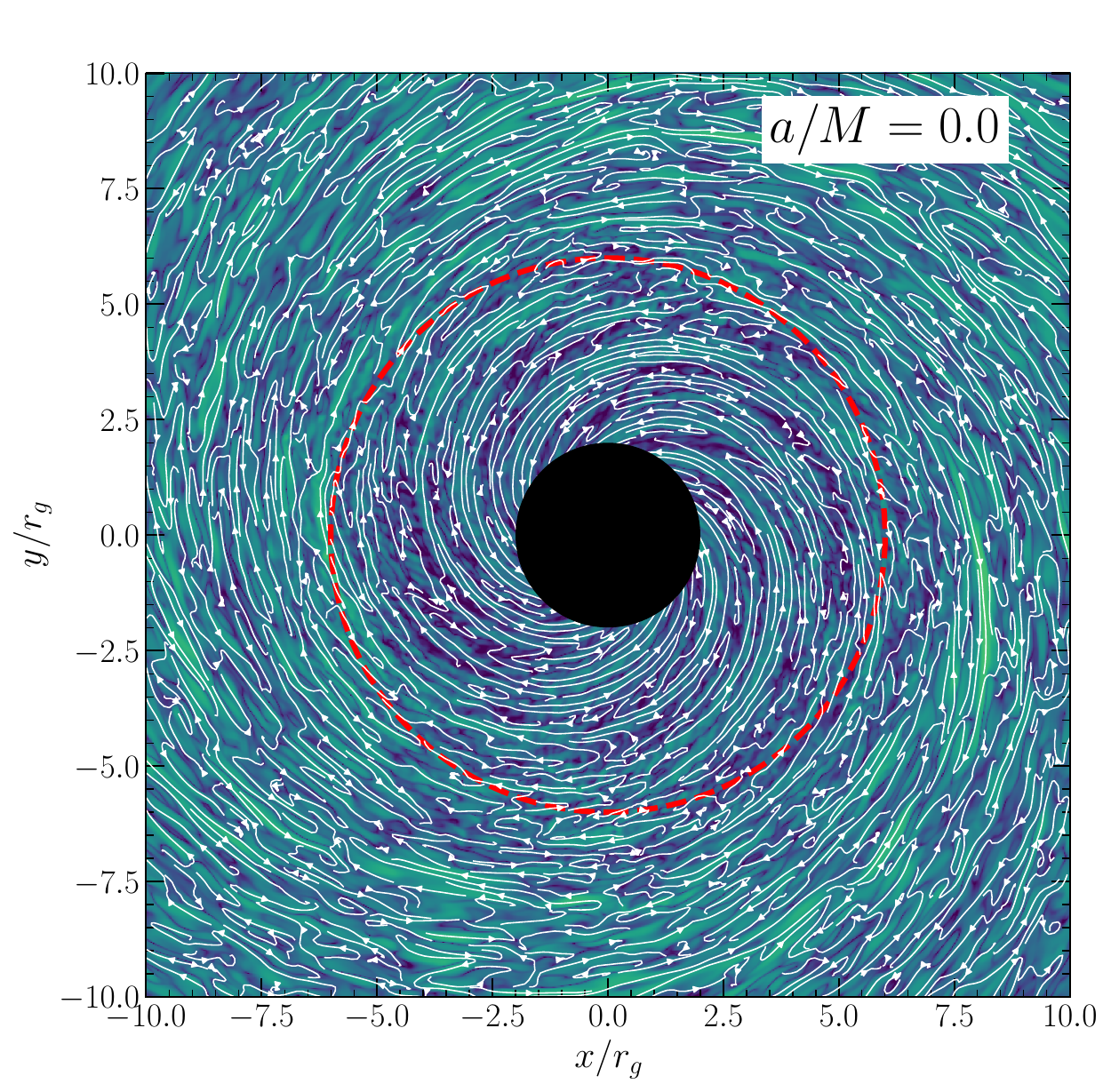}
    \includegraphics[width=0.33\linewidth]{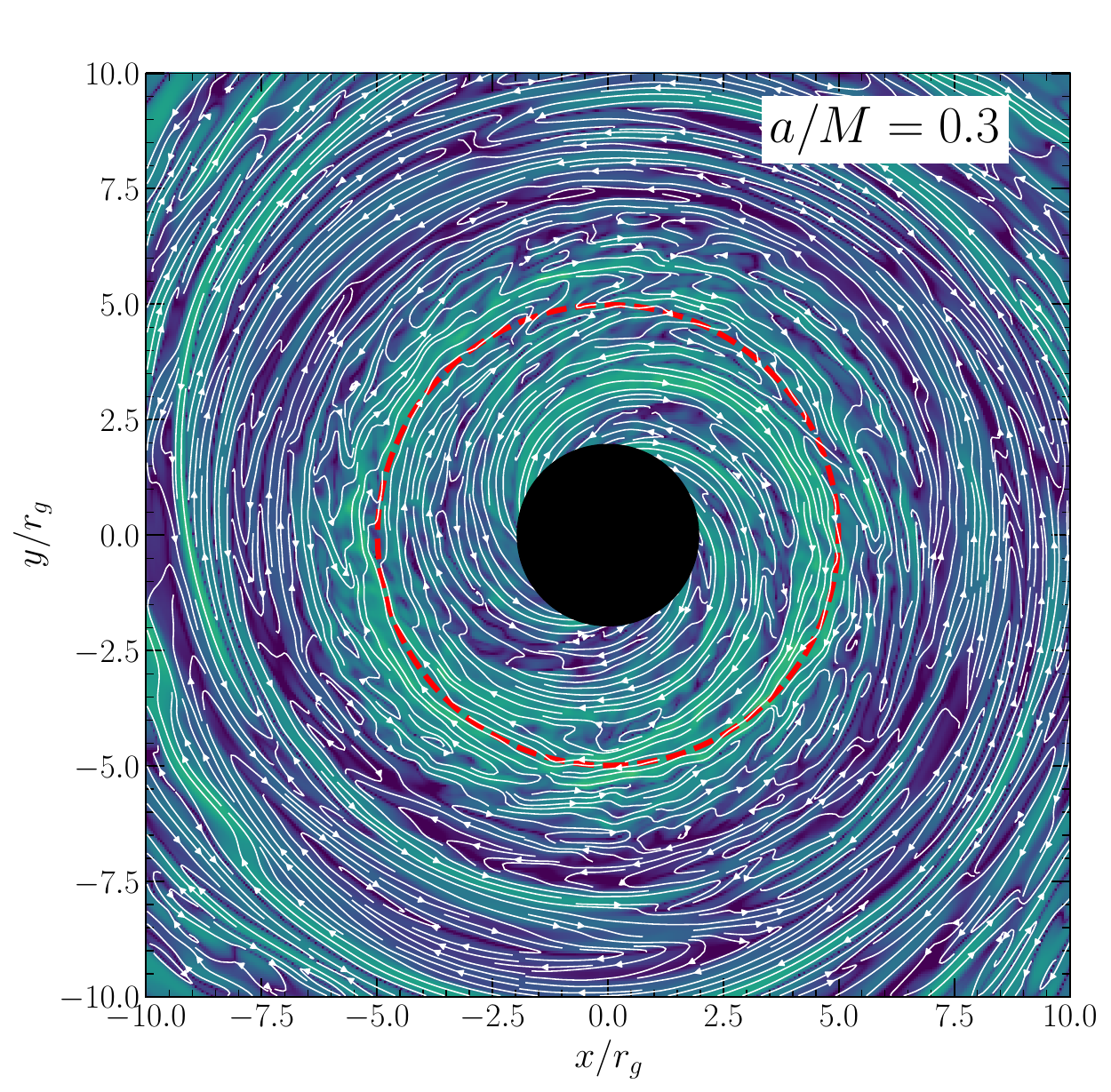}
    \includegraphics[width=0.33\linewidth]{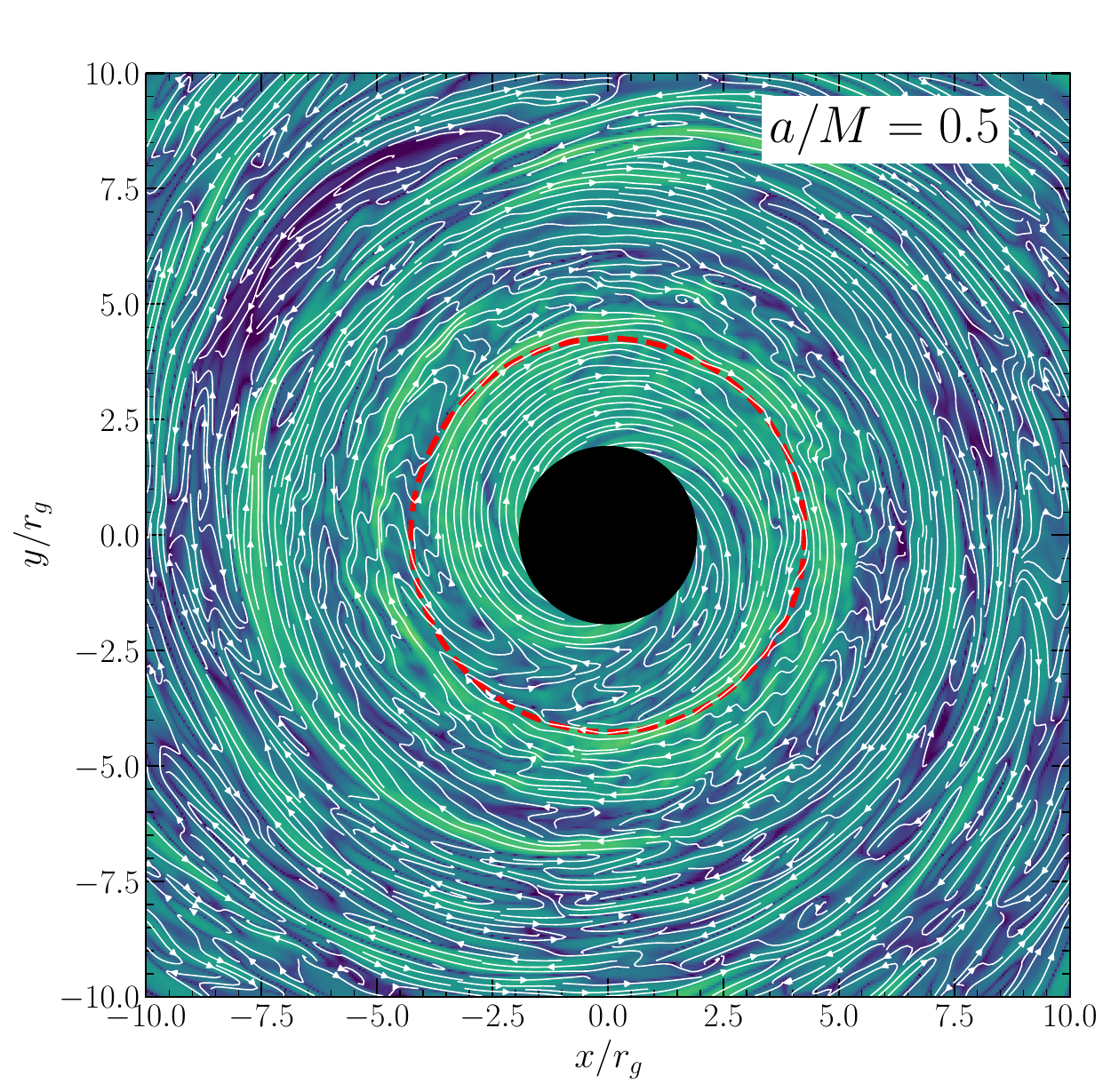}
    \includegraphics[width=0.33\linewidth]{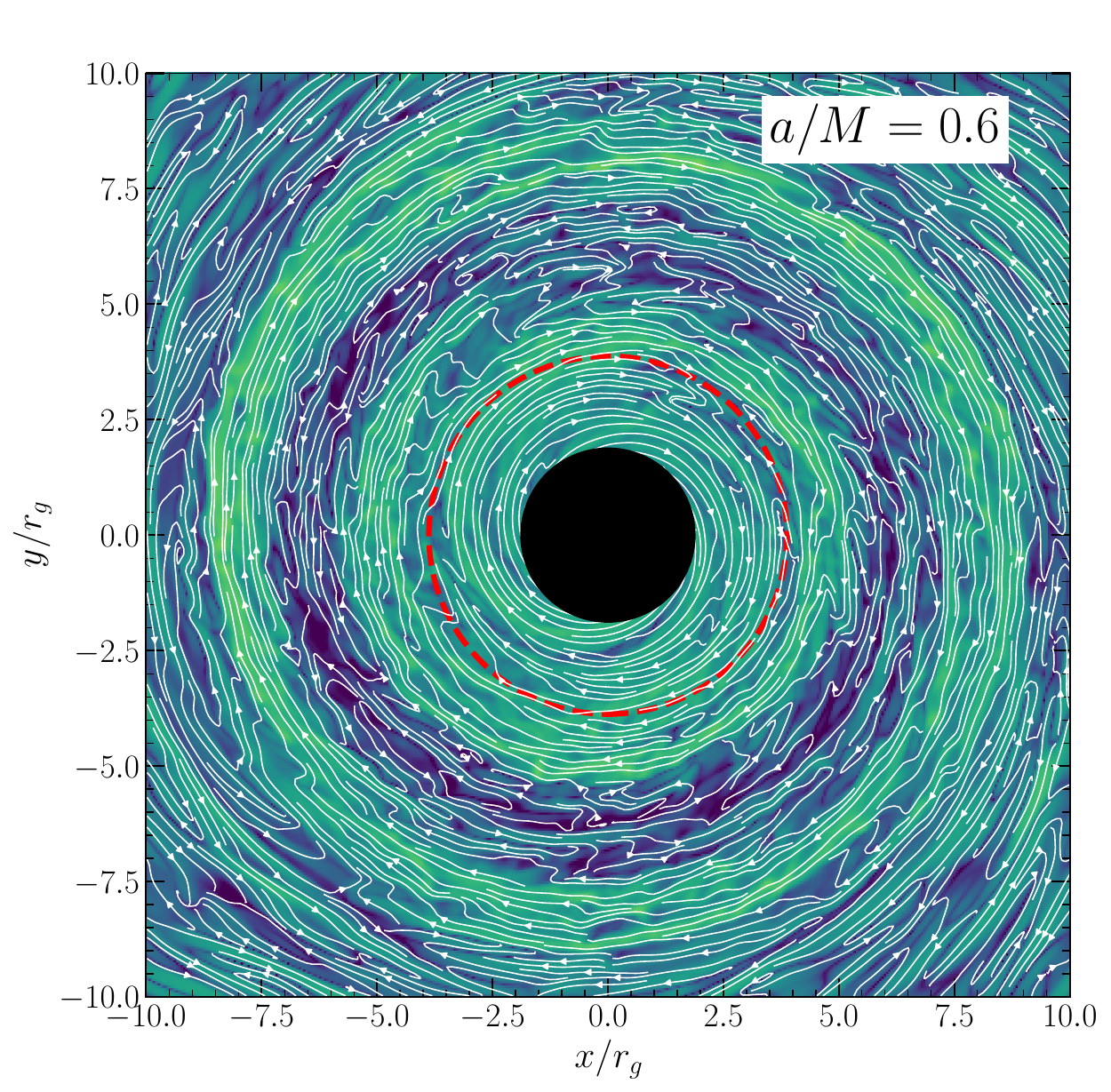}
    \includegraphics[width=0.33\linewidth]{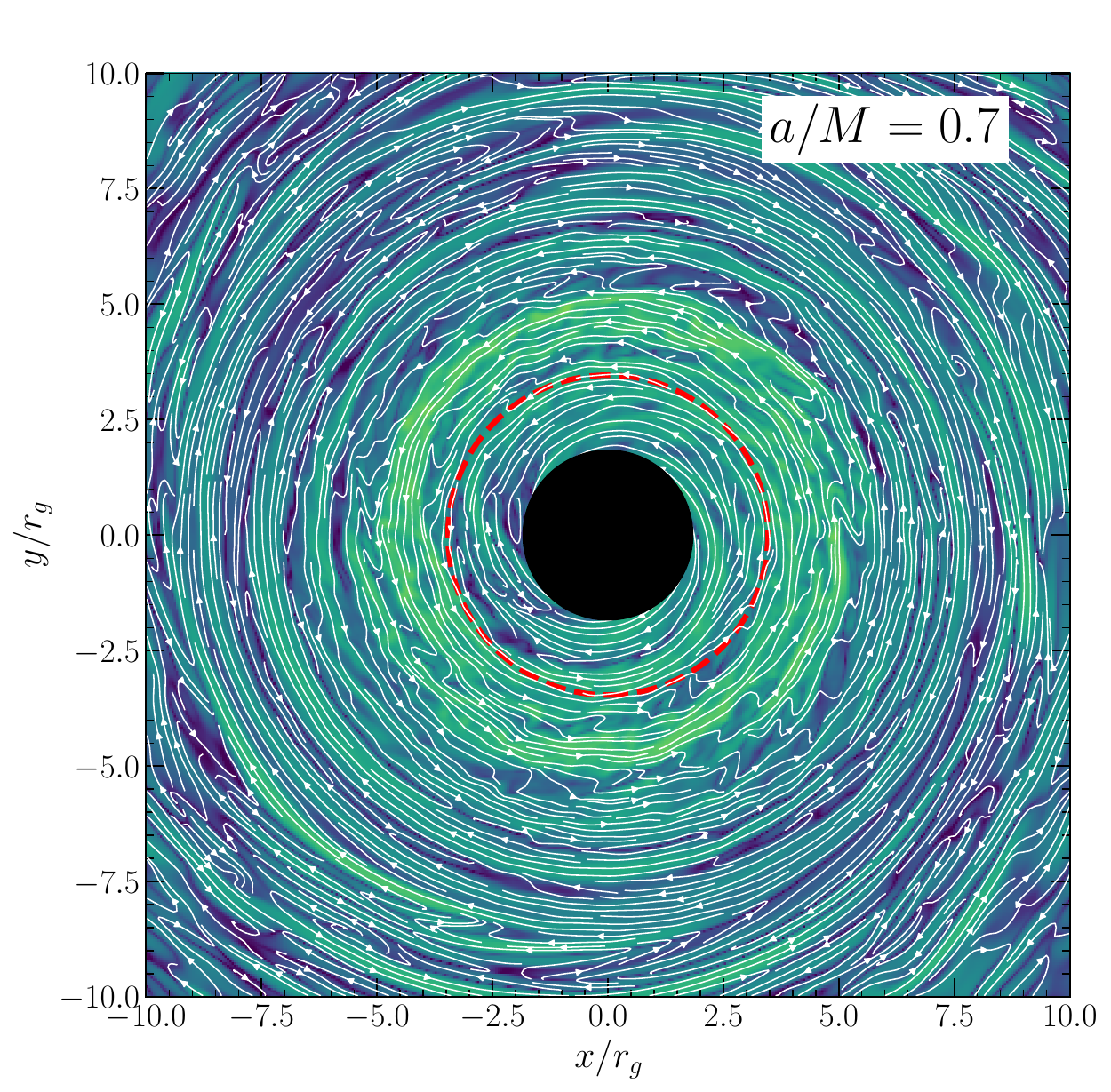}
    \includegraphics[width=0.33\linewidth]{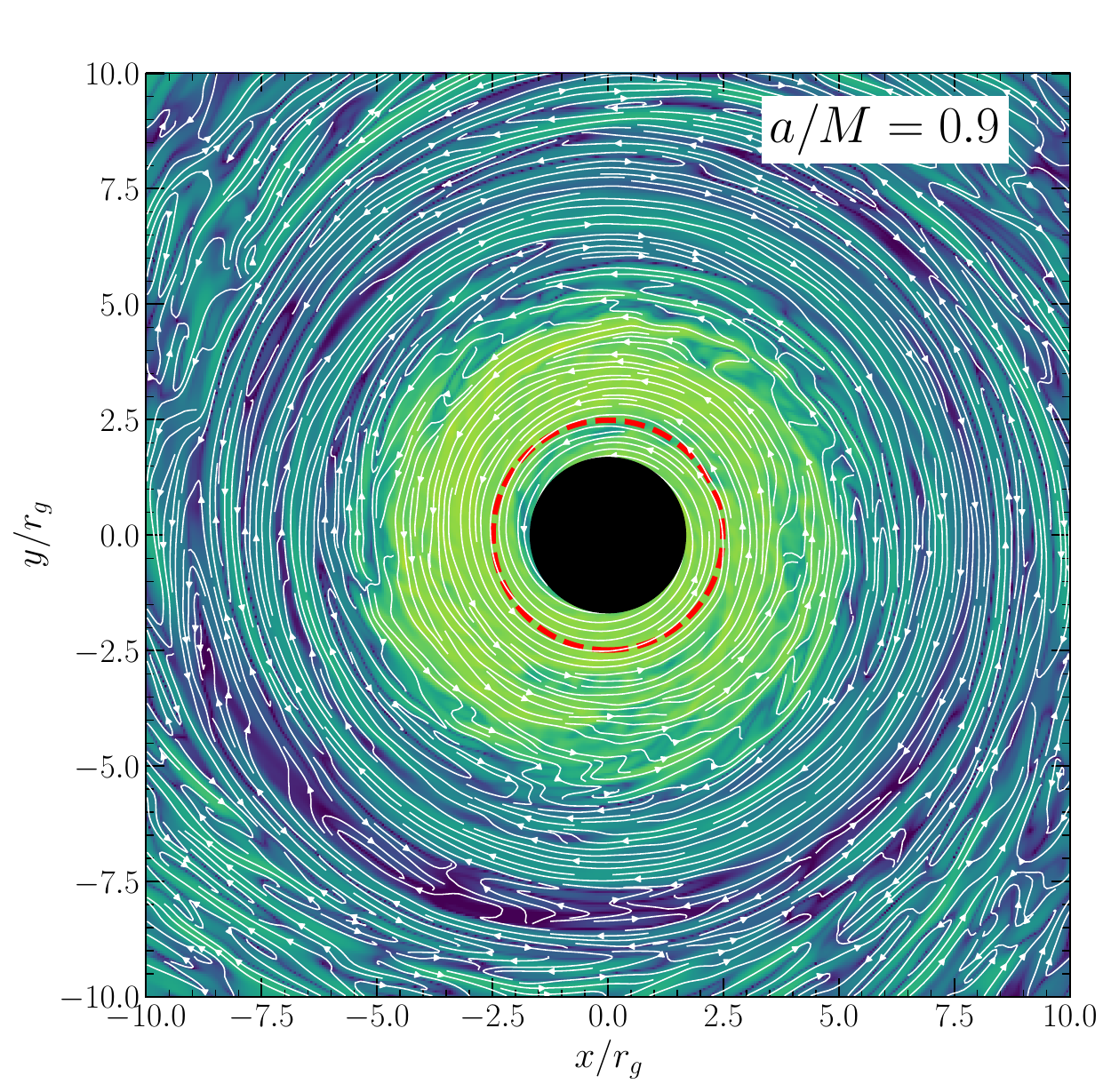}
    \caption{The magnetic field in the disc mid-plane ($z=0$) at the end-point of the simulations ($t/t_g=25,000$). The colour map shows the electromagnetic energy density ($u_\text{mag}$), whilst the white lines show spatial streamlines of the magnetic four-vector components, $b^x$ and $b^y$. It should be noted that these components do not, in general, satisfy the same divergence-free condition that the spatial magnetic field does. The thick dashed circular lines indicate the ISCO radius, whilst the solid black circles cover the interior of the outer horizon of the black hole. The spin of the black hole is labelled in the upper-right of each plot.
    }
    \label{fig:FaceMagneticFields}   
\end{figure*}

It is clear from Fig.\,\ref{fig:MagneticFields} that the magnitude of the $b^\phi$ component increases with increasing prograde spin throughout the plunging region. To demonstrate this clearly, in Fig.\,\ref{fig:FaceMagneticFields}, we plot a face-on $z=0$ slice of the magnetic field lines ($b^r$ and $b^\phi$) and the electromagnetic energy density for the final snapshot of each simulation. The magnetic fields both in the transition region just outside of the ISCO and in the plunging region itself appear both stronger and more toroidal with increasing prograde spin.
\par
We will first discuss the ordering of $b^\phi$ \emph{at the ISCO}. This is determined by the degree of field amplification in the transition region just exterior to the ISCO. As the location of the ISCO moves inward with increasing prograde spin, fluid elements may follow progressively more rapid stable circular orbits deeper into the potential well of the black hole. Thus, in the transition region, there is an ever stronger shear gradient. This shear gradient directly feeds the $b^\phi$ component of the magnetic field, transforming poloidal field into an enhanced toroidal field. Since fluid orbits are stable in this region (it is just exterior to the ISCO), there is ample time for this amplification to occur. This drives the strong spin ordering in the $b^\phi$ components as the fluid arrives \emph{at the ISCO}.
\par
Within the plunging region, to a very good approximation the magnetic fields are frozen-in to the geodesic inflow structure. Since this inflow has a considerable shear gradient, the $b^\phi$ component continues to grow throughout the plunging region. However, the timescale of the inward plunge is much shorter than the time spent traversing the ISCO transition region. Therefore, whilst $b^\phi$ continues to grow, the overall ordering in the magnitude of $b^\phi$ is closely maintained down to the horizon. In other words, in the low-spin and retrograde cases, the magnetic fields are not warped by the shear gradient in the deep potential of the black hole for enough time to catch up with the rapidly spinning prograde case, where this occurs on stable orbits, allowing time for many windings. 
\par
In addition to the toroidal fields, radial magnetic fields ($b^r$, see the top-left of Fig.\,\ref{fig:MagneticFields}) also contribute to the stress and must be considered\footnote{It should be noted that, in our coordinate system, $b^r$ can be an important contribution to $b_\phi$, particularly for near-extremal and retrograde spins.}. It is clear that this component is also strongly ordered by the prograde spin \emph{at the ISCO}. Indeed, this trend neatly aligns with the shear-driven strengthening of the toroidal fields if, for each spin, the fluid just exterior to the ISCO reaches a similar non-linear saturation state of the magnetorotational instability \citep[MRI, e.g.][]{balbusPowerfulLocalShear1991,balbusInstabilityTurbulenceEnhanced1998}. Turbulent radial motions will \emph{dynamically} regenerate $b^r$, which is then kinematically sheared into $b^\phi$. It is therefore only natural that a rise in the magnitude of the radial fields will accompany the shear-amplified strengthening of the toroidal fields, as the prograde spin is increased. 
\par
As fluid crosses into the plunging region, it undergoes considerable radial acceleration (see the top-left of Fig.\,\ref{fig:uprofiles}). In \citetalias{rulePlungingRegionThin2025a}, we found that this radial inflow is vertically stratified; the radial velocity increases sharply moving away from the equatorial plane. This vertical shear will feed the radial field ($b^r$) in the atmosphere of the plunging flow. Furthermore, the flow converges towards the equatorial plane, as the density and scale height drop due to rapid inward acceleration. This compression will quickly amplify the frozen-in radial field in the equatorial plane (see the top-left of Fig.\,\ref{fig:MagneticFields}). Since the equatorial ISCO crossing velocity increases with the prograde spin (see Section \ref{sec:geoinflow}), and since we expect the lightly loaded inflow far from the disc mid-plane to be very rapidly inflowing for all spins, we reason that the vertical shear will \emph{weaken} as the prograde spin is increased. Moreover, as the prograde spin increases, the plunging region shrinks, so there is less time for vertical compression to occur. These factors conspire to explain why $b^r$ rises most sharply in the plunging region for the \emph{retrograde} simulations in the top-left of Fig.\,\ref{fig:MagneticFields}.
\par
Nonetheless, despite this slight counteracting trend, it is clear from Fig.\,\ref{fig:MagneticFields} as a whole that there is an overwhelming strengthening of the magnetic fields, and therefore the magnetic stresses, as the prograde spin is increased. As flow crosses from the main body of the disc into the plunging region, there is a transition from a turbulent flow to a quasi-laminar flow. The magnetic fields undergo a similar transition; the turbulent field is advected and stretched by the coherent plunging flow into large-scale ordered spiral structures (see Fig.\,\ref{fig:FaceMagneticFields}). Since the radial gradient of the angular velocity is negative, the toroidal field that grows from the shearing flow is oppositely signed to the radial field that seeds it. Since the fields in the plunging region are ordered, this correlation is clearly visible (the spiral arms are inward-oriented in Fig.\,\ref{fig:FaceMagneticFields}). Finally, since the relevant stress is given by ${T_\mathrm{mag}}^r_\phi = - b^r b_\phi$, this negative correlation produces a positive Maxwell stress, ensuring that angular momentum is transported \emph{outward} (see the bottom-left of Fig.\,\ref{fig:MagneticFields}).
\section{Flux-freezing in the plunging region}
\label{sec:Flux-Freezing}
In Section \ref{sec:ISCOStress}, we analysed how the magnetic stress in the plunging region depends upon the black hole's spin. As the fluid transitions from a turbulent flow in the disc body to a quasi-laminar post-ISCO plunging flow, it is flux freezing that ensures a distinct transformation from a disordered, turbulent magnetic field in the main body of the disc, to an ordered field that is either aligned or anti-aligned with the spiral flow lines of the plunging region (Fig.\,\ref{fig:FaceMagneticFields}). In this section, we will develop this framework more carefully by solving the covariant induction equation of ideal GRMHD for a steady, axisymmetric system. This takes the following form:
\begin{equation}
    \label{eq:induction_eq}
    \nabla_\mu \left( b^\nu U^\mu - b^\mu U^\nu \right) = 0 \, , 
\end{equation}
where we have introduced the covariant derivative operator $\nabla_\mu$ and used both the magnetic four-vector $b^\mu$ and the four-velocity $U^\mu$. This equation may, however, be simplified to coordinate derivatives since it is the covariant divergence of an antisymmetric tensor. Using axisymmetry ($\partial_\varphi=0$) and adopting a steady state ($\partial_t=0$), we obtain:
\begin{equation}
    \partial_R \left( \sqrt{\vert g \vert} ( b^\nu U^R - b^R U^\nu ) \right) + \partial_z \left( \sqrt{\vert g \vert} ( b^\nu U^z - b^z U^\nu ) \right)= 0 \,.
\end{equation}
For convenience, we have adopted a \emph{cylindrical} Kerr-Schild coordinate system $(t,R,\varphi,z)$ in this section to facilitate vertical integration (see Appendix \ref{sec:CylindricalCoordAppendix}). To solve this equation, we adopt the following simple ansatz for the vertical structure: $b^\mu = \tilde{b}^\mu(R) e^{-z^2/H^2}$ and $U^\mu = \tilde{U}^\mu(R)$ \footnote{It should be noted that this is an approximation. We find the flow to be vertically stratified in our simulations.}, where $H \equiv H(R)$ is a function of the cylindrical radius that describes the scale height of the plunging fluid. Substituting this ansatz and integrating over $z$ we obtain:
\begin{equation*}
    \partial_R \left( \sqrt{\vert g \vert} \int_{-\infty}^{\infty} ( \tilde{b}^\nu \tilde{U}^R - \tilde{b}^R \tilde{U}^\nu ) e^{-z^2/H^2} \mathrm{d}z \right)  = 0 \, ,
\end{equation*}
where we have dropped the term arising from the $z$ derivative, since the height-integrated boundary terms vanish under our ansatz. We therefore find
\begin{equation}
    \label{eq:vert_avd_ind_eq}
    \partial_R \left( \sqrt{\vert g \vert} H ( \tilde{b}^\nu \tilde{U}^R - \tilde{b}^R \tilde{U}^\nu )\right)  = 0 \, .
\end{equation}
\par
From this point onwards, we will drop the $\tilde {X} $ notation and work exclusively with mid-plane quantities after vertical integration. When $\nu = R$, the left-hand side of Eq.\,\ref{eq:vert_avd_ind_eq} vanishes identically and offers no further information. We may also make use of $b^\mu U_\mu = 0$, which follows directly from the definition of $b^\mu$. For simplicity, we assume that $b^z \ll b^R, b^\varphi, b^0$, and shall ignore $b^z$ in the analysis.  We will also invoke the \citetalias{mummeryAccretionInnermostStable2023} assumption that the flow dynamics are determined entirely by gravity; that is, for the present problem $U^\mu$ is pre-specified and the magnetic field is passively advected. Under these assumptions, we may write down a system of three equations with three unknowns ($b^R, b^\varphi, b^0$) that can be solved explicitly:
\begin{align}
    R H \left( b^\varphi U^R - b^R U^\varphi \right) = C_\varphi \, ,\\
    R H \left( b^0 U^R - b^R U^0 \right) = C_0 \, , \\ 
    b^0 U_0 + b^\varphi U_\varphi + b^R U_R = 0 \, ,
\end{align}
where we have used $\sqrt{\vert g \vert} = R$ in cylindrical Kerr-Schild coordinates. Finally, to compare these solutions directly to our simulated profiles, we must transform the former to \emph{spherical} Kerr-Schild coordinates. Applying the transformations given in Appendix \ref{sec:CylindricalCoordAppendix}, and rewriting the unknown constants $C_\varphi$ and $C_0$ in terms of the magnetic four-vector components at the ISCO, $b^r_I$, $b^\phi_I$ (in \emph{spherical} Kerr-Schild coordinates!), and the scale height function evaluated at the ISCO $H_I \equiv H(r_I)$, we find:
\begin{align}
    \label{eq:brmodel}
    b^r &= \frac{r_I H_I}{r H} b^r_I \, ,\\
    \label{eq:bphmodel}
    b^\phi &= \frac{r_I H_I}{r H U^r} \left[b^r_I (U^\phi - U^\phi_I) + b^\phi_I U^r_I \right] \, , \\
    \label{eq:b0model}
    b^0 &= \frac{r_I H_I}{r H (-\gamma) U^r} \left[b^r_I (-\gamma U^0 + J U^\phi_I + 1) - b^\phi_I J U^r_I \right] \, .
\end{align}
It is also possible to express all of these expressions in the following general form:
\begin{equation}
    b^\mu = \frac{r_I H_I}{r H U^r} \left[ b^r_I U^\mu + {F_I^\star}^{\mu r} \right],
\end{equation}
where we have used the electromagnetic dual tensor evaluated at the ISCO:
\begin{equation}
    {F_I^\star}^{\mu \nu} = b_I^\mu U_I^\nu - b_I^\nu U_I^\mu.
\end{equation}
In deriving these expressions, we have used the fact that for a geodesic flow, $J=U_\phi=U_\varphi$ and $\gamma = - U_0$ are constants of motion and are fixed to their circular orbit values at the ISCO. Both $U^\phi_I$ and $U^r_I$ are their respective four-velocity components evaluated at the ISCO. Just as for the thermodynamic profiles, we must invoke a non-zero $U^r_I$ to avoid divergent behaviour at the ISCO. We therefore adopt the same constant offset geodesic model as \citetalias{mummeryAccretionInnermostStable2023}, treating $U^r_I$ as a free parameter. Finally, $U^\phi_I = U^\phi_{\rm geo}(r_I)$ is the azimuthal velocity of a circular orbit at the ISCO in spherical Kerr-Schild coordinates (see Eq.\,\ref{eq:uphigeo}).
\par
We assume that $\vert U^r_I \vert \ll 1$, i.e. much less than the speed of light ($c=1$), which is the scale for the other four-velocity components at $r_I$. Adding the constant offset $U^r_I$ to the radial four-velocity requires corresponding corrections to the time and azimuthal components in order to satisfy $U^\mu U_\mu=-1$. However, these corrections are small compared with $U^0_{\rm geo}$ and $U^\phi_{\rm geo}$ everywhere in the post-ISCO flow and may therefore be safely neglected. By contrast, $U^r$ consists entirely of $U^r_I$ at $r_I$ since $U^r_{\rm geo}(r_I)=0$.  Since we neglect these corrections, the four-velocity is only approximately normalised, with
\begin{equation}
    U^\mu U_\mu +1 = U^r_I \left[ g_{rr}(2U^r_{\rm geo} + {U^r_I}) +2g_{r0} U^0_{\rm geo}  + 2g_{r \phi} U^\phi_{\rm geo} \right].
\end{equation}
The quantity inside the square brackets is of order unity or smaller because the geodesic velocity scale is the speed of light ($c=1$) and the metric factors are $\mathcal{O}(1)$. Thus, $\vert U^r_I \vert \ll 1$ implies $\vert U^\mu U_\mu +1 \vert \ll 1$. In forthcoming work, we will include the additional corrections required to enforce $U^\mu U_\mu = -1$, although these do not substantively change our current findings.
\begin{figure*}
    \includegraphics[width=0.49\linewidth]{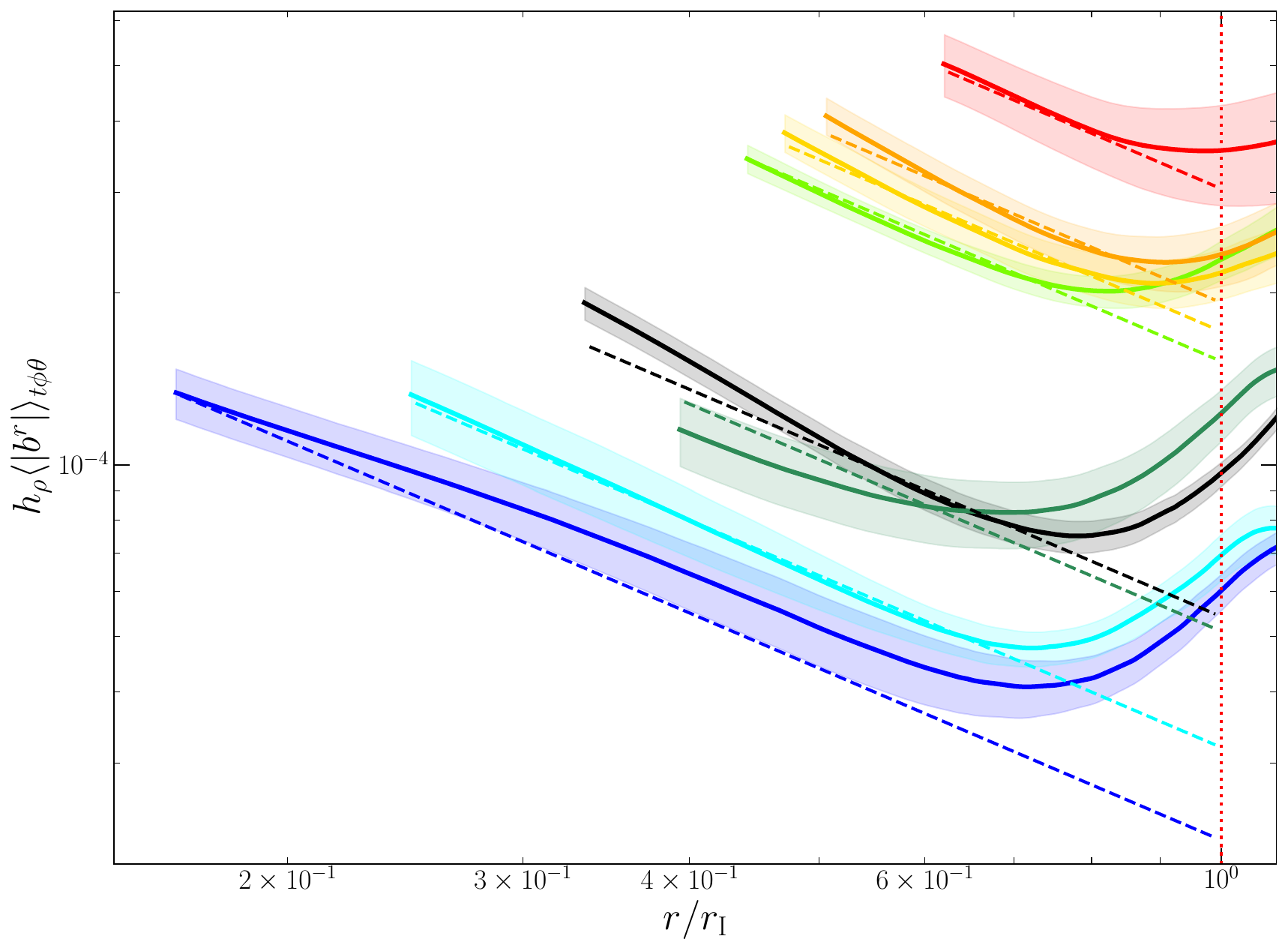}
    \includegraphics[width=0.49\linewidth]{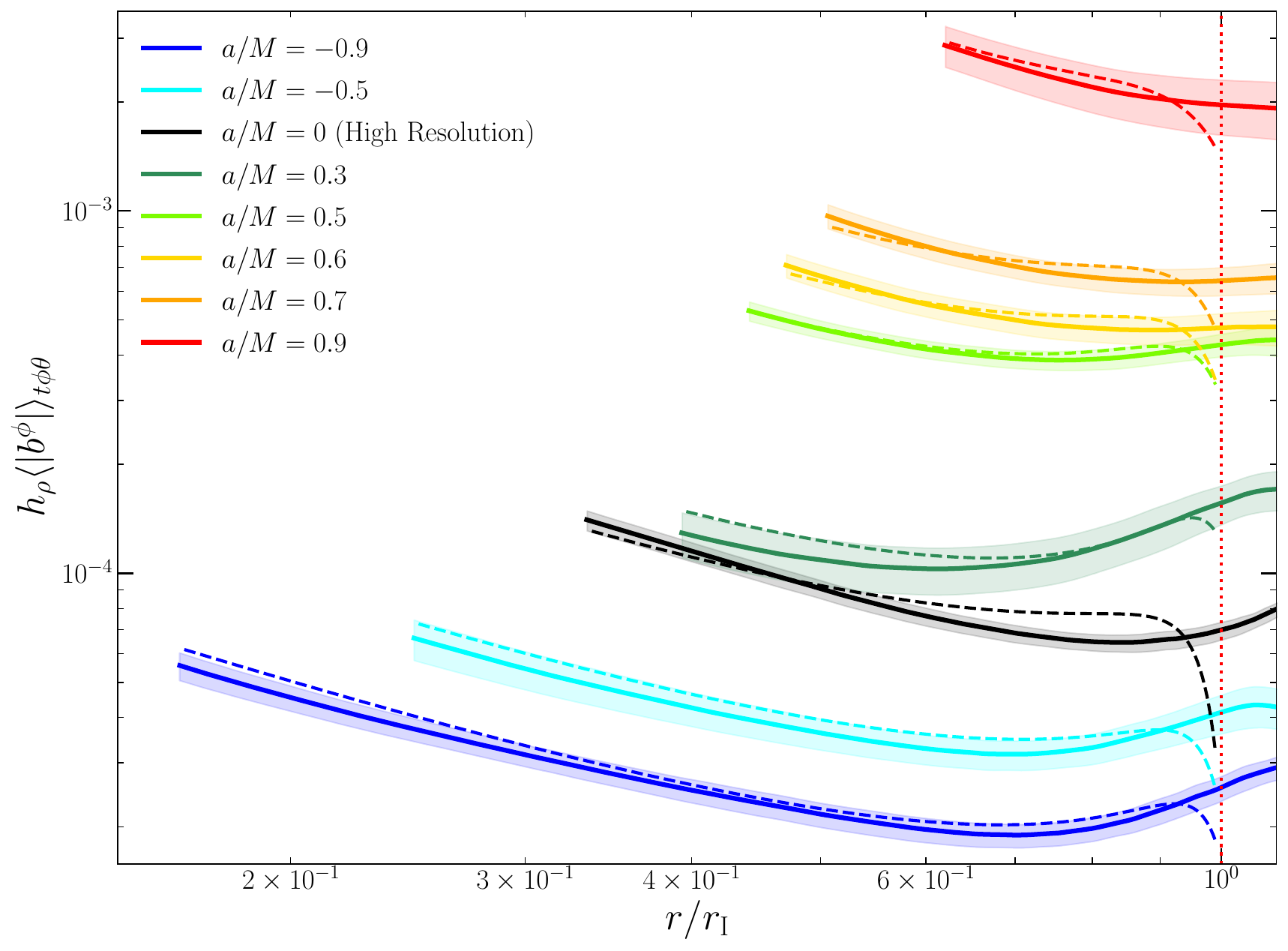}
    \includegraphics[width=0.49\linewidth]{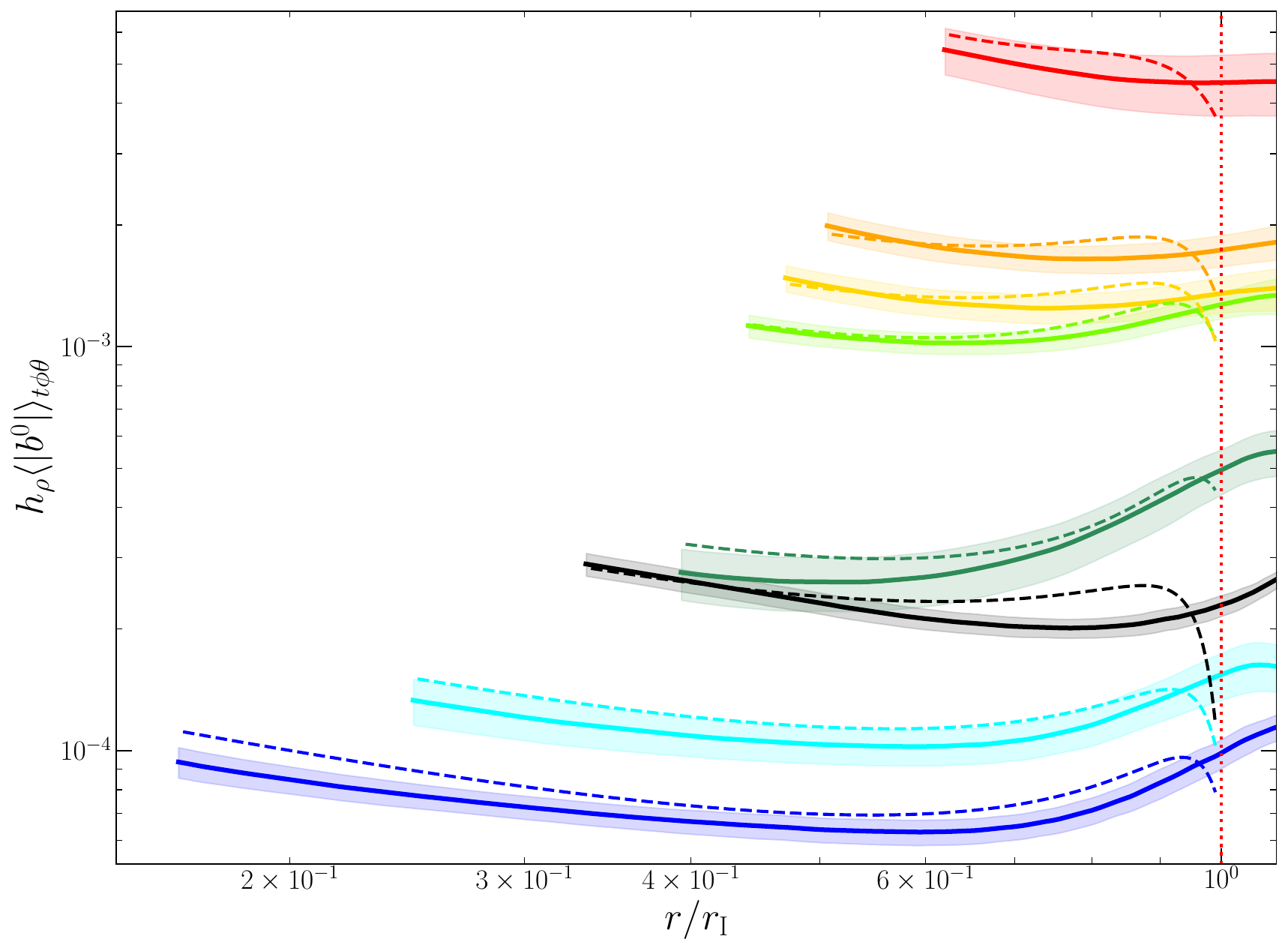}
    \caption{The components of the magnetic four-vector, $b^\mu$, multiplied by the scale height, $h_\rho$, as a function of radius. Clockwise from the top-left: $h_\rho b^r$, $h_\rho b^\phi$ and $h_\rho b^0$. The solid lines indicate the temporally, vertically and azimuthally averaged simulated profiles. The dashed lines are the flux-freezing models (Eqs. \ref{eq:brmodel}-\ref{eq:b0model}). We have simultaneously fitted the free parameters $b^r_I$ and $b^\phi_I$ to the simulated $b^r$ and $b^\phi$ profiles. In each case, we have fixed the offset velocity, $U^r_I$, using the associated simulated radial velocity profile.}
    \label{fig:flux_freeze_model}
\end{figure*}
\par
In Fig.\,\ref{fig:flux_freeze_model}, we plot components of the magnetic 4-vector (absolute values, then averaged as in Fig.\,\ref{fig:MagneticFields}) multiplied by $h_\rho$, which is the scale height measured by the following density weighted average of the polar angle from the mid-plane (in spherical Kerr-Schild coordinates):
\begin{equation}
    h_\rho = \frac{r}{t_f-t_i} \int_{t_i}^{t_f} \left[ \frac{\int^{2\pi}_{0}\int^{\theta_l}_{\theta_u} \vert \theta -\pi/2\vert \rho \sqrt{g_{[t \phi \theta]}} \mathrm{d}\theta \mathrm{d \phi}}{\int^{2\pi}_{0} \int^{\theta_l}_{\theta_u} \rho \sqrt{g_{[t \phi \theta]}} \mathrm{d}\theta \mathrm{d \phi}} \right] \mathrm{d}t \, ,
\end{equation}
where $g_{[t \phi \theta]}$ is the absolute value of the determinant of the $t \phi \theta$ ``submetric'', and $t_i, t_f, \theta_u, \theta_l$ are the boundaries for our usual averaging region (see \citetalias{rulePlungingRegionThin2025a}).
\par
The dashed lines are the flux-freezing model derived in this section (Eqs.\,\ref{eq:brmodel}-\ref{eq:b0model}). To set the free parameters of the model for comparison with each simulation, $U^r_I$ is fixed to the value of the simulated $U^r$ profile at the ISCO (Fig.\,\ref{fig:uprofiles}). We then fit for the remaining free parameters, $H_I b^r_I$ and $H_I b^\phi_I$, by minimising the total squared distance of the models from both the simulated $h_\rho b^r$ and $h_\rho b^\phi$ profiles simultaneously. To simplify the fitting process, we have not included the $b^0$ profile, as (in principle) it is fully specified by the other profiles and so is redundant information.
\par
Overall, this simple flux-freezing model does quite well in capturing the behaviour of the magnetic fields in the plunging region. Just as for the thermodynamic profiles, it is unsurprisingly least accurate when close to the ISCO. This is primarily because the dynamical offset geodesic model is not very accurate near to the ISCO. It is therefore unsurprising that our magnetic-field solutions for a prescribed flow $U^\mu$ suffer in the same region.
\par
On the other hand, Eq.\,\ref{eq:brmodel} predicts that $Hb^r$ should follow a simple geometric $1/r$ (power-law) dependence, with no dependence on the four-velocity $U^\mu$ at all. To some extent, this is an artefact of our simple equatorial model. Had we not neglected the vertical magnetic fields and modelled the vertical stratification of the flow, this would not be the case. However, since in our case the vertical fields are rather weak, we expect that the vertical shear will only modestly affect the radial field. In all cases, we observe that the simulated $Hb^r$ profile initially \emph{dips} (going inwards from the ISCO to the horizon) and then rises with a more-or-less constant slope, which is consistent with power-law dependence on logarithmic axes. We suggest that this dip could be due to \emph{non-ideal} MHD dissipation destroying poloidal field at the grid-scale of the simulation. This interpretation accords with our previous conviction that numerical magnetic dissipation is the cause of the considerable non-adiabatic heating that we have observed (see Fig.\,\ref{fig:kprofiles} in Section \ref{sec:MB23Test} or \citetalias{rulePlungingRegionThin2025a} for a detailed discussion).
\par
These shortcomings notwithstanding, it is gratifying that a relatively simple analytic solution continues to describe the averaged radial behaviour of the plunging flow in our 3D GRMHD simulations rather well. The ideal MHD treatment of these magnetic fields is an advancement of the existing \citetalias{mummeryAccretionInnermostStable2023} framework.
\par
With our equations for $b^r$, $b^\phi$ and $b^0$, we may now compute the \emph{covariant} component $b_\phi$:
\begin{equation}
    b_\phi = \frac{r_I H_I}{r H U^r} \left[ U^r_I {b_I}_\phi + (g_{\phi \mu}(r) - g_{\phi \mu}(r_I))(b_I^\mu U_I^r - b_I^r U_I^\mu)\right] \, ,
\end{equation}
where $g_{\mu \nu}(r)$ is understood to be the spherical Kerr-Schild metric evaluated at the mid-plane ($\theta = \pi/2$) at the radial coordinate $r$. Multiplying this by our expression for $b^r$, we may obtain the magnetic stress term ${T_\mathrm{mag}}^r_\phi = -b^rb_\phi$:
\begin{multline}
     -b^rb_\phi = -\frac{1}{U^r} \left(\frac{r_I H_I}{r H}\right)^2 \left[ U^r_I {b_I}_\phi b_I^r \right. \\
     + \left.(g_{\phi \mu}(r) - g_{\phi \mu}(r_I))(U_I^r b_I^\mu b_I^r - U_I^\mu (b_I^r)^2)\right] \, .   
\end{multline}
\par
Finally, we may now construct an explicit expression for the magnetic \cite{shakuraBlackHolesBinary1973} $\alpha$ parameter that was discussed in \citetalias{rulePlungingRegionThin2025a}:
\begin{equation}
    \label{eq:alphadef}
    \alpha \equiv \frac{T^{(\phi)(r)}}{P+b^2/2}.
\end{equation}
Once again, we define $T^{(\phi)(r)}$ to be the magnetic stress as measured in the local comoving frame of the fluid:
\begin{equation}
    T^{(\phi)(r)} = - e^{(\phi)}_\mu e^{(r)}_\nu  b^\mu b^\nu .
\end{equation}
To define this frame, we will adopt the tetrad defined in Appendix B of \cite{kulkarniMeasuringBlackHole2011} (as in \citetalias{rulePlungingRegionThin2025a}). The local basis covectors (evaluated in the disc mid-plane) are:
\begin{align}
    e^{(r)}_\mu = \sqrt{\frac{\Delta}{(1+U_r U^r)r^2}
    }\left( -\gamma U_r, \frac{r^2}{\Delta}(1+U_r U^r), 0, J U_r  \right) \, , \\
    e^{(\phi)}_\mu = \sqrt{\frac{\Delta}{1+U_r U^r}}\left( -U^\phi, 0, 0, U^0  \right) \, .
\end{align}
This tetrad is now defined to be spatially aligned with respect to the \emph{Boyer-Lindquist} global coordinate system \citep{boyerMaximalAnalyticExtension1967}, so the components of the four-velocity $U^\mu$ in these expressions are evaluated in those coordinates. Additionally, $\Delta \equiv r^2 - 2Mr + a^2$. To evaluate the local stress, we require the components of the magnetic four-vector $b^\mu$ in Boyer-Lindquist coordinates. Conveniently, our earlier solutions (Eqs.\,\ref{eq:brmodel}-\ref{eq:b0model}) are invariant to the transformation between spherical Kerr-Schild coordinates and Boyer-Lindquist coordinates \footnote{Mixing between the $\mu$ and $r$ coordinates under the transformation produces ${F^\star}^{rr}$ contributions, which vanish identically since ${F^\star}$ is an antisymmetric tensor.}. We may therefore replace each component in the expression with its Boyer-Lindquist counterpart, keeping the functional form intact. After some simplification, we obtain:
\begin{equation}
    T^{(\phi)(r)} = \frac{-rb^r}{1+U_r U^r}(b^\phi U^0 - b^0 U^\phi) \, .
\end{equation}
Upon substitution of our flux-freezing expressions for $b^\mu$, we find:
\begin{equation}
    \label{eq:comovingstress}
    T^{(\phi)(r)} = \frac{-rb_I^r}{(1+U_r U^r)U^r}\left( \frac{r_I H_I}{rH} \right)^2 \left[ {F_I^\star}^{\phi r} U^0 - {F_I^\star}^{0 r} U^\phi \right]  \, .
\end{equation}
Once again, it should be noted that the right-hand side of this expression is in \emph{Boyer-Lindquist} coordinates. 
\par
Finally, to complete our derivation for $\alpha$, we need to model both the gas pressure $P$ and the magnetic pressure $b^2/2 \equiv b^\mu b_\mu/2$. Adapting results from \citetalias{mummeryAccretionInnermostStable2023}, we find for the gas pressure $P$:
\begin{equation}
    \label{eq:gaspressure}
    P=P_I \left( \frac{K}{K_I} \right) \left(\frac{r_I H_I U_I^r}{r H U^r}\right)^{\gamma_\mathrm{ad}}, 
\end{equation}
where we have used mass conservation to determine $\rho$ and then used the expression $P=K\rho^{\gamma_\mathrm{ad}}$, keeping the $H$ and $K$ dependence explicit. Note that here ${\gamma_\mathrm{ad}}$ denotes the adiabatic index, \emph{not} $-U_0$. Furthermore, if we now contract our model for $b^\mu$ with itself, we find the following expression for the magnetic pressure $b^\mu b_\mu/2$ in Boyer-Lindquist coordinates:
\begin{multline}
    \label{eq:magpressure}
    \frac{b^\mu b_\mu}{2} = \frac{1}{2} \left(\frac{r_I H_I}{r H U^r}\right)^2 \left[(U^r_I)^2b^\mu_I {b_I}_\mu \right. \\
    +\left. \left(g_{\mu \nu}(r)-g_{\mu \nu}(r_I)\right) {F_I^\star}^{\mu r} {F_I^\star}^{\nu r} \right]\,.
\end{multline}
Substituting Eqs.\,\ref{eq:comovingstress}-\ref{eq:magpressure} into Eq.\,\ref{eq:alphadef} completes our analytic MHD model of the magnetic $\alpha$ parameter in the plunging region. 
\par
Examining Eqs.\,\ref{eq:comovingstress}-\ref{eq:magpressure}, it is clear that $\alpha$ will depend on the scale height $H$, unless ${\gamma_\mathrm{ad}}=2$.  This requires some additional modelling. While \citetalias{mummeryAccretionInnermostStable2023} have produced such a model for the case of gas pressure dominated flow ($P\gg b^2/2$), we find that this is not always the case in our GRMHD simulations. This is because, whilst both the gas pressure and the magnetic fields are sub-dominant to the gravitational forcing terms, there is nothing to stop the gas pressure and magnetic fields from being of comparable magnitude. It is therefore desirable to include magnetic pressure support (i.e. Eq.\,\ref{eq:magpressure}) in the scale height model adopted by \citetalias{mummeryAccretionInnermostStable2023} \cite[originally][]{abramowiczAccretionDisksKerr1997}\footnote{Note that we are able to ignore the effects of magnetic tension here because we have assumed that $b^\theta \approx 0$.}:
\begin{equation}
    H^2 = \frac{r^4(P+b^2/2)}{2GMr_I \rho} \,.
\end{equation}
Substituting in Eq.\,\ref{eq:gaspressure} for $P$, Eq.\,\ref{eq:magpressure} for $b^2/2$ and an expression for the density,
\begin{equation}
    \rho = \rho_I \left(\frac{r_I H_I U_I^r}{r H U^r}\right) ,
\end{equation}
we arrive at an implicit equation for $H$ that can be solved numerically. We note that the addition of magnetic support to the scale height model will also alter the \citetalias{mummeryAccretionInnermostStable2023} thermodynamic profiles. This analysis has the potential to be very interesting and will be explored in a future paper.
\par
\begin{figure}
    \centering
    \includegraphics[width=0.95\linewidth]{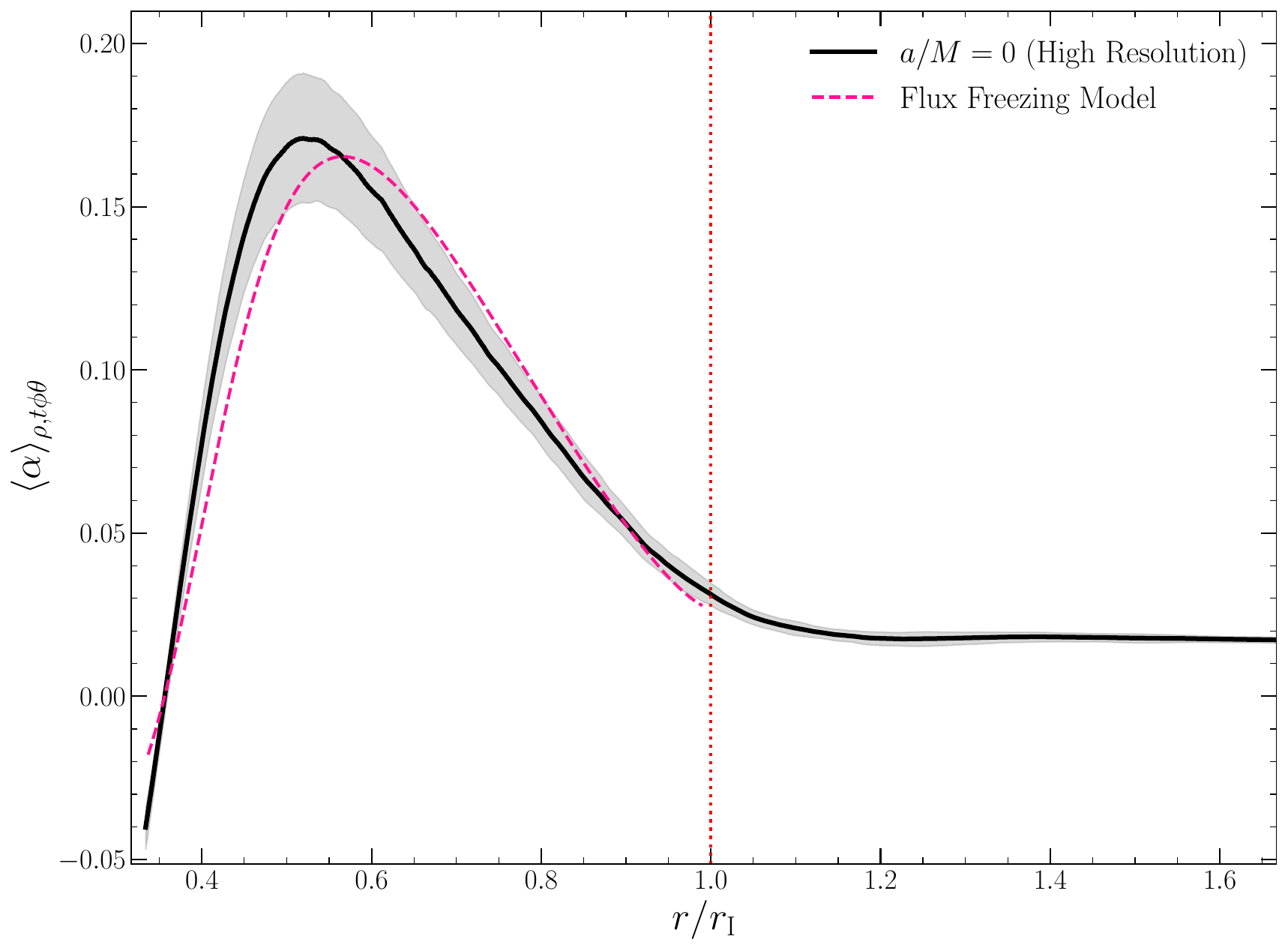}
    \caption{ The $\alpha$ parameter that we define in Eq.\,\ref{eq:alphadef}. The solid line is the vertical, azimuthal and temporal density weighted average $\alpha$ profile for our high-resolution Schwarzschild ($a/M=0.0$) simulation. The dashed line is the flux-freezing model that we have derived in Section \ref{sec:Flux-Freezing}. To determine the free parameters of the model, we have minimised the squared distance between the simulated profile and the model.}
    \label{fig:alpha_ff_model}
\end{figure}

With a model for $H$, it is possible to fully determine $\alpha(r)$ as a function of the following ISCO boundary parameters: $\beta_I = P_I/(b_I^2/2)$, $\tau_I = b^\phi_I/b_I^r$ and $U^r_I$. In Fig.\,\ref{fig:alpha_ff_model}, we plot the averaged $\alpha$ profile from the $a/M=0.0$ high-resolution Schwarzschild simulation. By minimising the squared distance between the simulated profile and the model, we are able to constrain the parameters of the model \footnote{To determine $K(r)$, we have once again used a simple power-law model, fixing the index to the best-fit parameter determined from the simulated $K$ profile.}. We over-plot the best-fit model with a dashed line. Overall, given the extensive modelling required, our analytic expression fits the simulated profile remarkably well. We clearly see the same distinctive sharp rise and subsequent fall off. Since the simulated magnetic fields are not \emph{exactly} frozen-in (due to non-ideal grid effects), we should not expect perfect agreement. Nonetheless, we believe that this analysis demonstrates that ideal MHD and flux-freezing account for the development of magnetic stresses in the plunging region \citep[][]{krolikMagnetizedAccretionMarginally1999}. Moreover, our modelling clearly shows that $\alpha$ is by no means constant in the plunging region; indeed, it loses its phenomenological character as a turbulent viscosity here. The comoving stress $T^{(\phi)(r)}$ and the total pressure $P+b^2/2$ are distinct quantities that evolve separately in this non-local, quasi-laminar regime (see \citetalias{rulePlungingRegionThin2025a} for a more detailed discussion).
\section{Breaking the spin-stress degeneracy?}
\label{sec:spinstressdegen}
From an observational perspective, the magnitude of electromagnetic stresses in the plunging region and the black hole spin are, to some extent, degenerate \citep[][]{gammieEfficiencyMagnetizedThin1999,mummeryRapidBlackHole2025a}. Both a high-spin, low-stress system and a low-spin, high-stress system will produce similar thermal spectra. Without any prior constraints on the relationship between the two quantities, they must be treated as individual free parameters. This introduces a degree of uncertainty to black hole spin estimates, even if the emission from the plunging gas and the dissipation from a finite ISCO stress are correctly accounted for.
\begin{figure}
    \centering
    \includegraphics[width=0.95\linewidth]{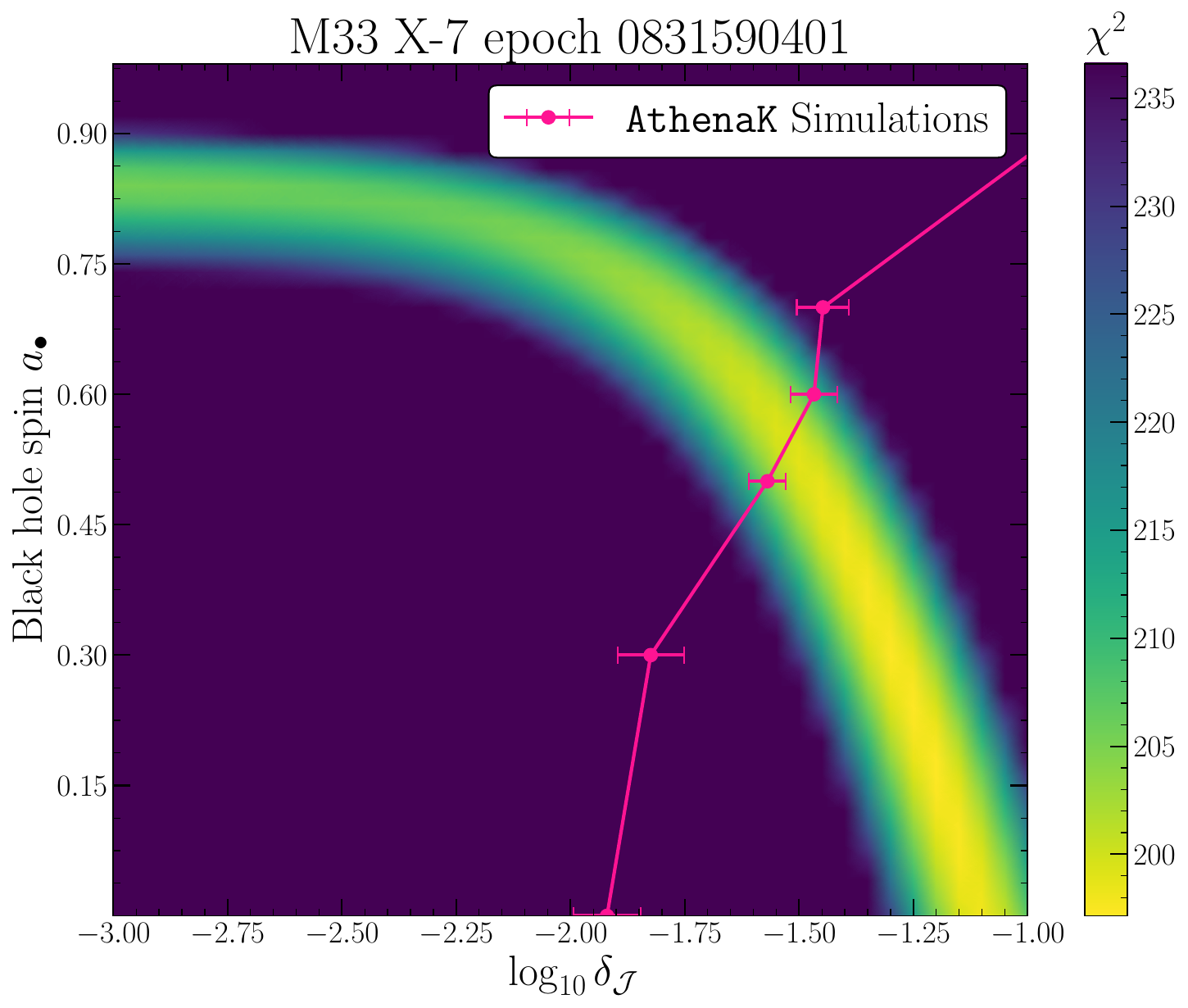}
    \caption{An adaptation of Figure 7 from \protect\cite{mummeryRapidBlackHole2025a}: the spin-stress degeneracy contour resulting from their analysis of an epoch of the X-ray spectrum of M33 X-7. The $\chi^2$ statistic that they find for each spin-stress pairing is shown by the colour bar, as a function of the black hole spin $a_\bullet$ and the stress quantified by $\delta_\mathcal{J}$. We over-plot the values of $\delta_\mathcal{J}$ that we have found for each of our {\tt ATHENAK} simulations, along with their associated uncertainties, across the black hole spin range \protect\citep[where they lie within the parameter domain specified by][]{mummeryRapidBlackHole2025a}. We connect the simulation data points with a line as a visual aid. There is a clear intersection between the spin-stress pairings that best fit the X-ray data (indicated by the lower values of $\chi^2$) and the connecting line. We caution that this does \emph{not} constitute a spin measurement; rather it is a demonstration that in principle the observational degeneracy between the spin and the stress is not fundamental.
    }
    \label{fig:spinstressdegen}
\end{figure}
\par
In this paper, we have found a strong coupling between the magnitude of the electromagnetic stresses in the plunging region and the spin of the central black hole across our suite of global GRMHD simulations that span the spin range. We have also identified a physical mechanism that is likely to be responsible for this trend in the physical regime that we have simulated. As we find that the magnitude of these stresses tends to increase with spin in the disc prograde direction\footnote{The increase in stress in the retrograde direction from $a/M=0.0$ to $a/M=-0.9$ is very slight and probably irrelevant given observational uncertainties.}, this suggests that it is, at least in principle, possible to break the observational degeneracy. To demonstrate this more clearly, in Fig.\,\ref{fig:spinstressdegen} we plot our computed spin-stress trend onto the $\chi^2$ distribution found by \cite{mummeryRapidBlackHole2025a}. This is a fit of the {\tt fullkerr} X-ray spectral model \citep[which includes emission from the plunging gas and the effects of a finite stress,][]{mummeryContinuumEmissionPlunging2024} to the X-ray spectrum of M33 X-7. It is to be emphasised that we do not claim this to be a \emph{measurement} of the spin of the black hole in M33 X-7. Rather, this is a demonstration that the trend in spin and stress that follows from our work crosses the `banana' shaped distribution of the spin-stress pairings that minimise the $\chi^2$ statistic for this particular dataset.
\par
Clearly, further work is required to transform this preliminary result into a more robust theoretical understanding of the spin-stress relationship that may be used for observational modelling. Our simulations only cover one particular initialisation of the magnetic fields. It is probable that in regimes with a greater magnetic flux onto the black hole, the physical mechanisms will alter considerably. A dynamically important magnetic field will obviously change our picture of a sub-dominant passive magnetic field, for instance (magnetised jets are common). In addition, our simulations adopt an ad-hoc cooling function to mimic radiative losses. A self-consistent treatment of radiation may well find the associated energetic losses to be dynamically important, particularly at higher accretion rates \citep[e.g.][]{zhangRadiationGRMHDModels2025,zhangRadiationGRMHDModels2026}. Nonetheless, the spin-stress coupling that we have found in our set of GRMHD simulations demonstrates that, at least in principle, the two quantities are not fundamentally degenerate.

\section{Conclusions}
\label{sec:conclusions}
In this paper, we have extended the work of \cite{rulePlungingRegionThin2025a} by comparing a set of simulations across the black hole spin range $a/M = \{-0.9,-0.5, 0.0,0.3,0.5,0.6,0.7,0.9\}$. We did this with two objectives in mind. Our first goal was to assess whether the analytic plunging solutions of \cite{mummeryAccretionInnermostStable2023} are supported by global GRMHD simulations outside of the Schwarzschild limit. The most important assumption of the \citetalias{mummeryAccretionInnermostStable2023} model is that gravity dominates the dynamics of the plunging flow. The model's assumptions require that the electromagnetic stresses transporting angular momentum from the plunging gas to the main body of the disc are at once large enough to avert an unphysical sharp thermodynamic transition at the ISCO, while also small enough that plunging, purely geodesic, trajectories nonetheless represent an excellent approximation to the flow dynamics. In Section \ref{sec:geoinflow}, we demonstrated that for all but the $a/M=0.9$ simulation, this is indeed the regime that our simulations fall into. In Section \ref{sec:MB23Test}, we then argued that, provided that non-adiabatic heating (probably due to magnetic dissipation, see \citetalias{rulePlungingRegionThin2025a}) is accounted for, we find good agreement between the \citetalias{mummeryAccretionInnermostStable2023} density solutions and the simulated radial profiles. This demonstrates that the other assumptions which underpin the \citetalias{mummeryAccretionInnermostStable2023} thermodynamic solutions are robust across the spin range of our thin, weakly magnetised (SANE) discs, which are designed to conform to the standard \cite{shakuraBlackHolesBinary1973} picture.
\par
Our second goal has been to study the behaviour of the electromagnetic stresses in the plunging region as a function of black hole spin. This is strongly motivated by the observational degeneracy between these quantities, as discussed in Section \ref{sec:spinstressdegen}. In Section \ref{sec:ISCOStress}, we found that there is a strong rise in the drop of (angle-averaged) specific angular momentum from the ISCO to the horizon as the spin is increased in the prograde direction. There is also a very weak rise in the same quantity as the spin is increased in the retrograde direction. These observations are explained by the fact that the outward magnetic flux of angular momentum strongly increases as a function of increasing prograde spin, whilst the physical size of the plunging region shrinks. Initially, from $a/M=-0.9$ to $a/M=0.0$, the shrinking of the plunging region reduces the accumulated drop in angular momentum. However, as the prograde spin is increased further (beyond $a/M=0.0$) the rise in the flux wins over, rapidly increasing the accumulated drop. It should be noted that this conclusion is subject to the caveat that we have only considered a set of weakly magnetised (SANE) simulations. It is possible that in the strongly magnetised (MAD) regime (e.g. \cite{narayanMagneticallyArrestedDisk2003}) this trend would change.
\par
In Section \ref{sec:ISCOStress}, we discussed how these electromagnetic stresses originate from the ordered correlation of azimuthal and radial components of the magnetic field that are frozen-in to the quasi-laminar plunging flow. These fields are amplified by the presence of a poloidal field as a source for toroidal field, engendered by the strong shear gradients in, and adjacent to, the plunging region. As the spin is increased in the prograde direction, the ISCO moves inwards, deeper into the potential well of the black hole. Magnetic fields may therefore directly tap into a stronger orbital shear gradient as accreting material crosses and flows into the plunging region. The stronger shear in turn produces stronger electromagnetic stresses, efficiently transporting angular momentum from the plunging fluid outwards into the disc.
\par 
To describe this process in detail, in Section \ref{sec:Flux-Freezing} we developed a new analytic framework to model the frozen-in magnetic fields in the plunging region. By solving the vertically integrated, ideal GRMHD induction equation for a steady, axisymmetric flow we were able to derive simple analytic functions for each component of the magnetic field (Eqs.\,\ref{eq:brmodel}-\ref{eq:b0model}). We assumed that the four-velocity is given by the \citetalias{mummeryAccretionInnermostStable2023} offset geodesic model and is unaltered by back-reaction from the magnetic fields. Once again, this assumes that gravity is the dominant forcing term in the relativistic Euler equation that determines the dynamical response.
\par
Our model clearly demonstrates the growth of toroidal field from poloidal field by the shear, and the organisation of a disordered field in the body of the disc into an ordered field by the coherent quasi-laminar plunging flow. Overall, we found good agreement when fitting these models to the simulated magnetic fields. We suggest that the discrepancies that arise between the two may be partly attributed to non-ideal MHD dissipation at the grid-scale of the simulation. This concords with our suggestion that this same dissipation is the cause of the non-adiabatic heating that we observe in the thermodynamic profiles (i.e. \citetalias{rulePlungingRegionThin2025a}).
\par
Our expanded model (i.e. \citetalias{mummeryAccretionInnermostStable2023} + flux-frozen fields) is quite similar to the model developed by \cite{gammieEfficiencyMagnetizedThin1999}, which also solves the ideal GRMHD radial transport problem for an equatorial, steady, axisymmetric plunging flow in the Kerr metric. The key difference is that \cite{gammieEfficiencyMagnetizedThin1999} \emph{does not} assume a fixed background flow $U^\mu$. Instead, they allow the electromagnetic stresses to self-consistently change the energy and angular momentum of the fluid as it plunges inwards. This comes at the expense of requiring a numerical solution to their system of non-linear algebraic equations, rather than having explicit analytic solutions. Whilst the \cite{gammieEfficiencyMagnetizedThin1999} model is in this sense more self-consistent, in the limit of weak electromagnetic stresses (when compared with gravitational forces), we have shown here that our geodesic modelling provides a good approximation to our full GRMHD simulations. An additional distinction is that our modelling includes a treatment of the thermodynamics, whilst the \cite{gammieEfficiencyMagnetizedThin1999} model is zero-temperature ($T=0$). We may therefore model the dependence of each quantity on the scale height, which we find to be crucial when making comparisons with the simulations.
\par
We also constructed a flux-freezing model for the formal \cite{shakuraBlackHolesBinary1973} $\alpha$ parameter (Eqs.\,\ref{eq:alphadef}, \ref{eq:comovingstress}-\ref{eq:magpressure}), which reproduced the characteristic rise and fall profile that we observe in our GRMHD simulations (see also \citetalias{rulePlungingRegionThin2025a} and \cite{lancovaRadiativeGRMHDSimulations2026}). This demonstrates that the comoving stress and the total pressure are distinct quantities that evolve separately in this region. We therefore argue that attempting to ascribe a local coupling to them via an $\alpha$ parameter has little physical significance. Our modelling is similar in spirit to the `mean field' component of the variable $\alpha(r)$ model proposed by \cite{pennaShakuraSunyaevViscosityPrescription2013}, which is itself based upon the \cite{gammieEfficiencyMagnetizedThin1999} model for the large-scale magnetic fields in the plunging region. \cite{abramowiczUniversalBehaviour$a$viscosity2026} have also recently proposed a different, more phenomenological model for $\alpha(r)$, including its behaviour within the plunging region, which they calibrate against simulated profiles.
\par
Finally, to construct an $\alpha$ model it was necessary to introduce magnetic pressure support to the existing scale height framework of \citetalias{mummeryAccretionInnermostStable2023} \citep[originally,][]{abramowiczAccretionDisksKerr1997}. Independently of its connection to $\alpha$, understanding the impact that magnetic fields may have on the scale height behaviour of the plunging region, and its connection to the thermodynamics (i.e.  \citetalias{mummeryAccretionInnermostStable2023}), will be the subject of future investigations.
\par
Overall, in this paper we have demonstrated that the dynamical and thermodynamic framework of \citetalias{mummeryAccretionInnermostStable2023} agrees amicably with our set of GRMHD simulations of thin, weakly magnetised accretion discs around black holes across the spin range. We further hope to have illuminated the physical processes that underpin the development of electromagnetic stresses in the plunging region that avert any discontinuous behaviour at the ISCO. Although these results are promising, it is clear that there is a wide parameter space of magnetic states \citep[e.g. the MAD regime][]{narayanMagneticallyArrestedDisk2003} to consider that may considerably alter our physical picture of the plunging region. We hope to explore this in future studies. Moreover, whilst computationally efficient, our ad-hoc cooling function is a potential shortcoming of our current simulations. Alternative cooling prescriptions may alter the plunging region profiles that we have presented in this paper. Other recent works \citep[e.g.][]{zhangRadiationGRMHDModels2025,zhangRadiationGRMHDModels2026,lancovaRadiativeGRMHDSimulations2026} include radiation self-consistently. It will be interesting to compare to these to determine what impact radiative forces may have on both the transport of angular momentum and more generally the flow dynamics inside and adjacent to the plunging region.

\section*{Acknowledgements}
The authors would like to thank the anonymous reviewer for their comments which greatly improved the paper. JR was supported by a Science and Technology Facilities Council studentship [grant number ST/Y509474/1]. A.M. acknowledges support from the Ambrose Monell Foundation, the W.M. Keck Foundation and the John N. Bahcall Fellowship Fund at the Institute for Advanced Study. An award for computer time was provided by the U.S. Department of Energy’s (DOE) Innovative and Novel Computational Impact on Theory and Experiment (INCITE) Program. This research used resources from the Argonne Leadership Computing Facility, a U.S. DOE Office of Science user facility at Argonne National Laboratory, which is supported by the Office of Science of the U.S. DOE under Contract No. DE-AC02-06CH11357. The authors would like to acknowledge the use of the University of Oxford Advanced Research Computing (ARC) facility in carrying out this work. http://dx.doi.org/10.5281/zenodo.22558
\section*{Data Availability}
The X-ray spectrum of M33 X-7 is publicly available. Numerical results will be shared upon reasonable request with the corresponding author.



\bibliographystyle{mnras}
\bibliography{Jake}




\appendix

\section{$U^\phi$ Geodesic Solution in SKS coordinates}
\label{sec:AppA}
Following Equation\,14 of \cite{mummeryInspiralsInnermostStable2022}, it is straightforward to construct $U^\phi$ in Boyer-Lindquist coordinates from the conserved angular momentum $U_\phi = J_I$ and the energy $U_0 = -\gamma_I$ of the ISCO orbit:
\begin{equation}
    U^\phi_{\mathrm{BL}} = \frac{2\gamma_Ia+J_I(r-2)}{r(r^2-2r+a^2)}.
\end{equation}
Note that we will set $M=1$ in this appendix. In Boyer-Lindquist coordinates, $U^\phi$ diverges at the horizon ($r=r_+$), since by definition $\Delta_+ = r_+^2-2r_++a^2=0$. To remove this coordinate singularity, we move to horizon-penetrating SKS coordinates\footnote{Since the $r$ coordinate is identical in both systems, we do not denote the coordinate system for $r$ itself or for $U^r$.}:
\begin{equation}
    U^\phi_{\mathrm{SKS}} = U^\phi_{\mathrm{BL}} + \frac{a U^r}{\Delta} = \frac{2\gamma_Ia+J_I(r-2)+arU^r}{r\Delta}.
\end{equation}
It is useful to write $U^r$ in terms of $J_I$ and $\gamma_I$ rather than in the condensed form of Eq.\,\ref{eq:urgeo},
\begin{equation}
    \begin{aligned}
    U^r = -\frac{1}{r^2}\sqrt{
    -\Delta r^2+(r^4+a^2r^2+2a^2r)\gamma_I^2 - 4aJ_I\gamma_Ir
     - r(r-2)J_I^2 
    }\,,\\
    = - \frac{1}{r^2} \sqrt{
        \Delta\left((\gamma_I^2-1)r^2+2\gamma_I^2r-J_I^2\right)+\left(2\gamma_Ir-aJ_I\right)^2
    }\,,
    \end{aligned}
\end{equation}
which may be verified directly from Equation\,10 of \cite{mummeryInspiralsInnermostStable2022}. Using this second expression for $U^r$ we may write $U^\phi_\mathrm{SKS}$ as:
\begin{equation}
    U^\phi_\mathrm{SKS} = \frac{a}{r^2\Delta}\left(A+\frac{\Delta J_I}{a}- \sqrt{B\Delta + A^2}
    \right).
\end{equation}
where
\begin{align*}
    A = 2 r \gamma_I - aJ_I, \\
    B = (\gamma_I^2 - 1)r^2 + 2 \gamma_I^2 r - J_I^2, \\
    \Delta = r^2 - 2r +a^2.
\end{align*}
At the horizon, where $\Delta_+=0$, this form is indeterminate. To eliminate this behaviour, we multiply both the numerator and denominator by the conjugate of the numerator, giving us our final expression from Eq.\,\ref{eq:uphigeo}:
\begin{equation}
    U^\phi_\mathrm{SKS} = \frac{\Delta J_I^2+2 a J_I A - a^2 B}{r^2\left( a A + \Delta J_I + a\sqrt{B\Delta + A^2}\right)}.
\end{equation}

\section{Cylindrical Kerr-Schild Coordinates}
\label{sec:CylindricalCoordAppendix}
In Section \ref{sec:Flux-Freezing}, it was necessary to vertically integrate the induction equation to account for the role of the scale height in the flux-freezing dynamics. To perform this integration globally, rather than in the close vicinity of the disc mid-plane (as we do for the simulated quantities), it is highly advantageous for the vertical distance from the mid-plane to be explicit in the coordinate system. It is therefore natural to define a \emph{Cylindrical} Kerr-Schild coordinate system. To do this, we will start with the well known Cartesian Kerr-Schild coordinates $(t,x,y,z)$ \citep[][]{kerrNewClassVacuum1965}, where the invariant line element is:
\begin{multline}
    \label{eq:inv_line_el_cks}
    \mathrm{d}s^2  = -\mathrm{d}t^2 + \mathrm{d}x^2 + \mathrm{d}y^2 +\mathrm{d}z^2 \\ +\frac{2Mr^3}{r^4+a^2z^2} \left[ \mathrm{d}t + \frac{r(x\mathrm{d}x + y\mathrm{d}y)-a(x\mathrm{d}y - y\mathrm{d}x)}{r^2+a^2} + \frac{z\mathrm{d}z}{r} \right]^2 \, ,
\end{multline}
where $r$ is the usual (spherical) radial coordinate, which may be expressed in terms of $x,y,z$:
\begin{equation}
    \label{eq:radial_coord_def_cks}
    r^2 = \frac{x^2+y^2+z^2-a^2}{2} + \sqrt{\left( \frac{x^2+y^2+z^2-a^2}{2} \right)^2 +a^2 z^2}.
\end{equation}
We will now define the following simple transformation to a Cylindrical coordinate system $(t,R,\varphi,z)$ where the $t$ and $z$ coordinates are left unchanged from their Cartesian Kerr-Schild definitions:
\begin{align}
    x = R \cos(\varphi), \\
    y = R \sin(\varphi).
\end{align}
We may deduce the following 1-form transformations between the Cylindrical and Cartesian coordinates:
\begin{align}
    \mathrm{d}x = \cos(\varphi) \mathrm{d}R - R \sin(\varphi) \mathrm{d}\varphi, \\
    \mathrm{d}y = \sin(\varphi) \mathrm{d}R + R \cos(\varphi) \mathrm{d}\varphi.
\end{align}
Substituting these into the expression for the invariant line element Eq.\,\ref{eq:inv_line_el_cks}, we obtain:
\begin{multline}
    \mathrm{d}s^2  = -\mathrm{d}t^2 + \mathrm{d}R^2 + R^2\mathrm{d}\varphi^2 +\mathrm{d}z^2 \\ +\frac{2Mr^3}{r^4+a^2z^2} \left[ \mathrm{d}t + \frac{rR\mathrm{d}R-aR^2\mathrm{d}\varphi}{r^2+a^2} + \frac{z\mathrm{d}z}{r} \right]^2 \, .
\end{multline}
\par
Finally, we relate these cylindrical Kerr-Schild coordinates to the spherical Kerr-Schild coordinates that we have used extensively throughout this paper. First of all, it is important to distinguish the spherical radial coordinate $r$, from the cylindrical radial coordinate $R$. Indeed, since $x^2+y^2 = R^2$, we may relate the two via Eq.\,\ref{eq:radial_coord_def_cks}:
\begin{equation}
    r^2 = \frac{R^2+z^2-a^2}{2} + \sqrt{\left( \frac{R^2+z^2-a^2}{2} \right)^2 +a^2 z^2}.
\end{equation}
Note that in the mid-plane, where $z=0$, this simplifies to $r^2 = R^2 - a^2$. Additionally, using the relationship between Cartesian Kerr-Schild and Spherical Kerr-Schild coordinates,
\begin{align}
    x = \sqrt{r^2+a^2} \sin(\theta) \cos(\phi + \arctan(a/r)) , \\
    y = \sqrt{r^2+a^2} \sin(\theta) \sin(\phi + \arctan(a/r)) , \\
    z = r \cos(\theta),
\end{align}
we may also make the following set of identifications to relate the Cylindrical Kerr-Schild coordinates to Spherical Kerr-Schild coordinates:
\begin{align}
    R = \sqrt{r^2+a^2} \sin(\theta), \\
    \varphi = \phi + \arctan(a/r) , \\
    z = r \cos(\theta).
\end{align}
Note that the cylindrical azimuthal coordinate $\varphi$ is \emph{not} the same as the Spherical Kerr-Schild azimuthal coordinate $\phi$.
\par
Finally, we derive the following 1-form transformations between Spherical Kerr-Schild coordinates and Cylindrical Kerr-Schild coordinates:
\begin{align}
    \mathrm{d}R = \frac{r\sin(\theta)}{\sqrt{r^2+a^2}} \mathrm{d}r+\sqrt{r^2+a^2} \cos(\theta)\mathrm{d}\theta , \\
    \mathrm{d}\varphi = \mathrm{d}\phi - \frac{a}{r^2+a^2} \mathrm{d}r ,\\
    \mathrm{d}z = \cos(\theta) \mathrm{d}r - r\sin(\theta) \mathrm{d}\theta .
\end{align}

\bsp	
\label{lastpage}
\end{document}